\documentclass[fleqn,usenatbib]{mnras}
\usepackage{hyperref}
\usepackage{multirow}
\usepackage{graphicx}
\usepackage{times}
\usepackage{float}
\usepackage{pdflscape}
\usepackage{threeparttable}
\usepackage{amsmath}
\usepackage{amssymb}
\usepackage{xspace}
\usepackage{subcaption}
\usepackage{newtxtext}
\title[Clumpy disc-wind in MCG--03--58--007]{Evidence for a clumpy disc-wind in the star forming Seyfert\,2 galaxy MCG--03--58--007}
\author[Matzeu et al.]
{G. A. Matzeu$^{1,2}$\thanks{Correspondence to: gabriele.matzeu@esa.int}, 
V. Braito$^{2,3}$, J. N. Reeves$^{3}$, P. Severgnini$^4$, L. Ballo$^1$, A. Caccianiga$^4$,    
\newauthor
S. Campana$^4$, 
C. Cicone$^4$,
R. Della Ceca$^4$,
M. L. Parker$^{1}$,
M. Santos-Lle{\'o}$^{1}$,
and N. Schartel$^1$\\
$^{1}$European Space Agency (ESA), European Space Astronomy Centre (ESAC), E-28691 Villanueva de la Ca\~{n}ada, Madrid, Spain\\
$^2$INAF -- Osservatorio Astronomico di Brera, Via Bianchi 46, I-23807 Merate (LC), Italy\\
$^3$Center for Space Science and Technology, University of Maryland Baltimore County, 1000 Hilltop Circle, Baltimore, MD 21250, USA\\
$^4$INAF -- Osservatorio Astronomico di Brera, Via Brera 28, I-20121 Milano, Italy\\}
		\newcommand{\Msun}{\mbox{$M_{\odot}$}\xspace}
		\newcommand{\vunit}{\mbox{\,km\,s$^{-1}$}\xspace}
		
		\newcommand{\ang}{\mathring{\mathrm{A}}}
 		
		\newcommand{\pds}{PDS\,456\xspace}		
		\newcommand{\mcg}{MCG--03--58--007\xspace}		

		\newcommand{\feka}{Fe\,K$\alpha$\xspace}

		\newcommand{\fcov}{f_{\rm cov}\xspace}
		\newcommand{\fekb}{Fe\,K$\beta$\xspace}
		\newcommand{\fexxv}{Fe\,\textsc{xxv}\xspace}
		
		\newcommand{\fexxvi}{Fe\,\textsc{xxvi}\xspace}

		\newcommand{\mw}{\dot M_{\rm w}}

		\newcommand{\vturb}{v_{\rm turb}\xspace}

		\newcommand{\nh}{N_{\rm H}}
		\newcommand{\lognh}{\log(N_{\rm H}/\rm{cm}^{-2})}
		\newcommand{\logxi}{\log(\xi/\rm{erg\,cm\,s}^{-1})}

		\newcommand{\lbol}{L_{\rm bol}}

		\newcommand{\lion}{L_{\rm ion}}

		\newcommand{\mbh}{M_{\rm BH}}

		\newcommand{\fe}{Fe\,K\xspace}
        \newcommand{\iron}{iron\,K\xspace}
		\newcommand{\ergs}{\rm erg\,s^{-1}\xspace}

		\newcommand{\rg}{R_{\rm g}}
		\newcommand{\ev}{\rm eV}		
		\newcommand{\kev}{\rm keV}
		\newcommand{\los}{line-of-sight\xspace}
		\newcommand{\vw}{v_{\rm w}}
		\newcommand{\vwc}{v_{\rm w}/c}		
				
		\newcommand{\kms}{\rm km\,s^{-1}\xspace}

		\newcommand{\fig}{Fig.\,}

		\newcommand{\ks}{\rm ks}
		\newcommand{\chis}{\chi^{2}/\nu}
		\newcommand{\dchis}{\Delta\chi^{2}/\Delta\nu}
		\newcommand{\C}{C/\nu}
		\newcommand{\dC}{\Delta C/\Delta\nu}
		\newcommand{\knorm}{\rm ph\,cm^{-2}\,s^{-1}\,keV^{-1}}

		\newcommand{\suzaku}{\emph{Suzaku}\xspace} 
		\newcommand{\nustar}{\emph{NuSTAR}\xspace}		
		\newcommand{\xmm}{\emph{XMM-Newton}\xspace}

		\newcommand{\chandra}{\emph{Chandra}\xspace}

		\newcommand{\xmmnu}{\textit{XMM-Newton} \& \textit{NuSTAR}\xspace}

		\newcommand{\pexmon}{\texttt{pexmon}\xspace}

		\newcommand{\myt}{\texttt{MYTorus}\xspace}

		\newcommand{\mekal}{\texttt{mekal}\xspace}

		\newcommand{\xstar}{\textsc{xstar}\xspace}
		\newcommand{\xspec}{\textsc{xspec}\xspace}

		\newcommand{\rgsproc}{\textit{rgsproc}\xspace}
		\newcommand{\sas}{\textsc{sas}\xspace}
		
		\newcommand{\rgscombine}{\textit{rgscombine}\xspace}

\begin{document}

\date{\today}

\pagerange{\pageref{firstpage}--\pageref{}} \pubyear{?}

\maketitle
\label{firstpage}

\maketitle
\begin{abstract}

\noindent We report the results of a detailed analysis of a deep simultaneous $130\,\rm ks$ \xmmnu observation of the nearby ($z=0.0315$) and bright ($L_{\rm bol}\sim3\times10^{45}\ergs$) starburst-AGN Seyfert\,2 system: MCG--03--58--007. From the broadband fitting we show that most of the obscuration needs to be modeled with a toroidal type reprocessor such as \texttt{MYTorus} \citep{MurphyYaqoob09}. Nonetheless the signature of a powerful disc-wind is still apparent at higher energies and the observed rapid short-term X-ray spectral variability is more likely caused by a variable zone of highly ionized fast wind rather than by a neutral clumpy medium. We also detect X-ray emission from larger scale gas as seen from the presence of several soft narrow emission lines in the RGS, originating from a contribution of a weak star forming activity together with a dominant photoionized component from the AGN.

\end{abstract}
%\pagebreak

%\pagebreak

\section{introduction}

It is widely accepted that the central engine of active galactic nuclei (AGN) is powered by accretion of matter onto a super-massive black hole (SMBH). They are also thought to harbor circumnuclear material in a toroidal structure (i.e., dusty torus) that absorbs and reprocesses the high-energy radiation emitted from these central regions. Although the dusty torus is predicted to be ubiquitous in AGN by the unification scheme \citep{Antonucci93}, the exact geometry, location and composition are poorly understood. Nonetheless, important advances have been made in recent years through the observed long-term variability (i.e., months--years) of the X-ray absorber's column density ($\nh$). This ruled out the standard view of an homogeneous `doughnut' shaped absorber in favour of a more `clumpy' torus \citep[e.g.,][]{Risaliti02,Markowitz14} characterized by a distribution of a large number of individual clouds. 
   
Since the first detection of resonance \iron shell absorption lines blueshifted to rest-frame energies of $E>7\,\kev$ in luminous AGN \citep{Chartas03,Reeves03,Pounds03}, high velocity outflows have become an essential component in the overall understanding of AGN. The faster components of these winds are thought to occur as a result of the accretion process and originates in the inner regions near the black hole whereas the warm absorbers at lower velocities probably originate from much further out e.g., $>10\,\rm pc$ and greater \citep[e.g.,][]{Blustin05,Reeves13}. Fast disc-winds are present in $\sim40\%$ of the bright AGN \citep{Tombesi10,Gofford13}, suggesting that their geometry is characterized by a wide opening angle, as confirmed in the luminous quasar \pds by \citet{Nardini15}. These winds are considered the key ingredient of AGN feedback models and might represent the missing link in the observed galactic feedback process, by driving massive molecular outflows out to large ($\sim$kpc) scales in galaxies \citep{Cicone14,Tombesi15,Feruglio15,Cicone18b}. 

Disc-winds are characterized by a high column density ($N_{\rm H}\sim 10^{23}\,\rm cm^{-2}$) and a mean velocity $\left \langle \vw \right \rangle\sim0.1c$ \citep{Tombesi10}. These high velocities can result in a large amount of mechanical power, possibly exceeding the $0.5$--$5\%$ of the bolometric luminosity ($\lbol$), suggesting a significant AGN feedback contribution to the evolution of the host galaxy (\citealt{King03,KingP03,DiMatteo05,HopkinsElvis10,Tombesi15}). However, the velocity of the \fe absorbers can span over a wide range, from as low as a few $\times100-1000$\vunit (more typical of what is seen in the soft X-ray warm absorbers; e.g., \citealt{Kaastra00,Blustin05,Reeves13}) up to mildly relativistic values of $\sim0.2-0.4c$ in the most extreme cases (e.g., \citealt{Chartas02,Reeves09,Parker17,Parker18iras}. A recent simultaneous \xmmnu observation of the luminous quasar \pds carried out in March 2017 revealed a new relativistic component of the fast wind while observed in the low-flux state \citep{Reeves18b}, which provided the evidence of the multi-phase structure disc-wind. In recent studies a positive correlation was also found between the outflow velocity of the disc-wind and the X-ray luminosity in \pds \citep{Matzeu17b} and IRAS\,13224--3809 \citep{Pinto18}, which is what is expected in a radiatively driven wind scenario.

\mcg is a nearby ($z=0.0315$) bright Seyfert\,2 galaxy. It was first selected as a Compton-thick AGN (i.e., $\nh>10^{24}\,\rm cm^{-2}$) candidate due to its faint X-ray flux \citep{Severgnini12} as measured in a $10\,\rm ks$ observation with \xmm in 2005. A longer $80\,\rm ks$ \suzaku observation in 2010 revealed an obscured AGN but in the Compton-thin (i.e., $\nh\sim10^{23}\,\rm cm^{-2}$) regime with two deep blue-shifted absorption troughs between $7.5$--$8.5\,\kev$ \citep[][B18 hereafter]{Braito18}. These are associated with two zones of a highly ionized ($\logxi\sim5.5$) disc-wind with column density of $\lognh\sim23.7$ and $\lognh\sim23.9$, and outflowing at a mildly relativistic speed of $\vw\sim0.1$--$0.2c$. A more recent $130\,\rm ks$ simultaneous \xmmnu observation confirmed the presence of the wind and showed a rapid $\nh$ variation on a timescale of $\sim1\,\rm day$ (B18). Such short-term variability suggests the presence of inhomogeneous material as part of the disc-wind, rapidly crossing the line-of-sight. The work presented in B18 was focused on the analysis of the hard X-ray band and on the detection of the blueshifted iron\,K absorption features in \mcg. 

The main motivation of this follow up paper is to undertake a full broadband analysis of \mcg with the simultaneous \xmmnu data by parameterizing the observation with a more physically motivated model for the neutral absorber such as \myt \citep{MurphyYaqoob09}. With this comprehensive broadband analysis we can: (a) investigate whether the spectral variability, as seen through the hardening of the spectra, is to a change in the clumpy neutral absorber or to an inhomogeneous highly ionized disc-wind; (b) to investigate in detail, for the first time in this source, the physical properties of the soft X-ray emission. This paper is organized as follows: in Section\,\ref{sec:Observations and Data Reduction}, we summarize the data reduction, whereas in Section\,\ref{sec:Soft X-rays Spectral Analysis} we describe the spectral analysis that focuses on the RGS spectra. In Section\,\ref{subsec:The broadband xmmnu analysis} we focus on the broadband spectral analysis of the simultaneous \xmmnu observation of \mcg which includes a time-dependent spectral analysis. In Section\,\ref{sec:Discussion and conclusion} we discuss the possible origin of the observed soft X-ray emission lines and we discuss whether the obscuration event is caused by a transiting neutral absorber or by an inhomogeneous highly ionized disc-wind. In Section\,\ref{sec:Discussion and conclusion} we also discuss the energetic and location of the highly ionized absorber and compare it with what was found in B18. 

In this paper the values of $H_0=70$\,km\,s$^{-1}$\,Mpc$^{-1}$ and $\Omega_{\Lambda_{0}}=0.73$ are assumed throughout and errors are quoted at the $90\%$ confidence level ($\Delta \chi^{2}=2.71$ and $\Delta C=2.71$) for one parameter of interest unless otherwise stated.

\begin{table*}
%\begin{minipage}{100mm}

\centering

\begin{tabular}{cccccc}
\hline

                        &\multicolumn{3}{c}{\xmm}                                                                         &\multicolumn{2}{c}{\nustar}\\
\hline   
Obs.~ID                                                 &\multicolumn{3}{c}{$0764010101$}                                 &\multicolumn{2}{c}{$60101027002$}\\
%\\
Instrument                      &PN                        &MOS\,1$+$MOS\,2            &RGS\,1$+$RGS\,2                   &FPMA                        &FPMB\\                   
%\\
Start Date, Time (UT)           &2015-12-06, 12:56:06      &2015-12-06, 12:33:16       &2015-12-06, 13:33:07              &2015-12-06, 10:36:08        &2015-12-06, 10:36:08\\
%\\
End Date, Time (UT)             &2015-12-08, 01:25:11      &2015-12-08, 01:29:12       &2015-12-08, 01:33:25              &2015-12-09, 17:21:08        &2015-12-09, 17:21:08\\
%\\
Duration(ks)                    &131.3                     &258.0                      &266.4                             &281.8                       &282.8\\
%\\
Exposure(ks)$^{\rm a}$          &59.9                      &195.3                      &213.1                             &131.4                       &132.0\\
\\
Flux$_{(0.5-2)\rm keV}^{b}$     &0.0877                    &0.0879                      &0.0927                            &--                          & --\\
%\\
Flux$_{(2-10)\rm keV}^{b}$      &1.95                      &1.94                       &--                                &1.57                        &1.63\\
%\\
Flux$_{(3-50)\rm keV}^{b}$      &--                        &--                         &--                                &4.93                        &5.12\\
%\\
\hline
\end{tabular}
%\end{minipage}
\caption{Summary of the 2015 simultaneous \xmmnu observation of \mcg.}
\vspace{-5mm}
\begin{threeparttable}
\begin{tablenotes}
	\item[a] Net Exposure time, after background screening and dead-time correction.
	\item[b] Observed fluxes in the 0.5--2\,keV, 2--10\,keV and 3--50\,keV bands in units $\times10^{-12}$\,erg cm$^{-2}$ s$^{-1}$.	
\end{tablenotes}
\end{threeparttable}
\label{tab:summary_obs}
\end{table*}
%\pagebreak

  \begin{figure}
  \includegraphics[width=0.45\textwidth]{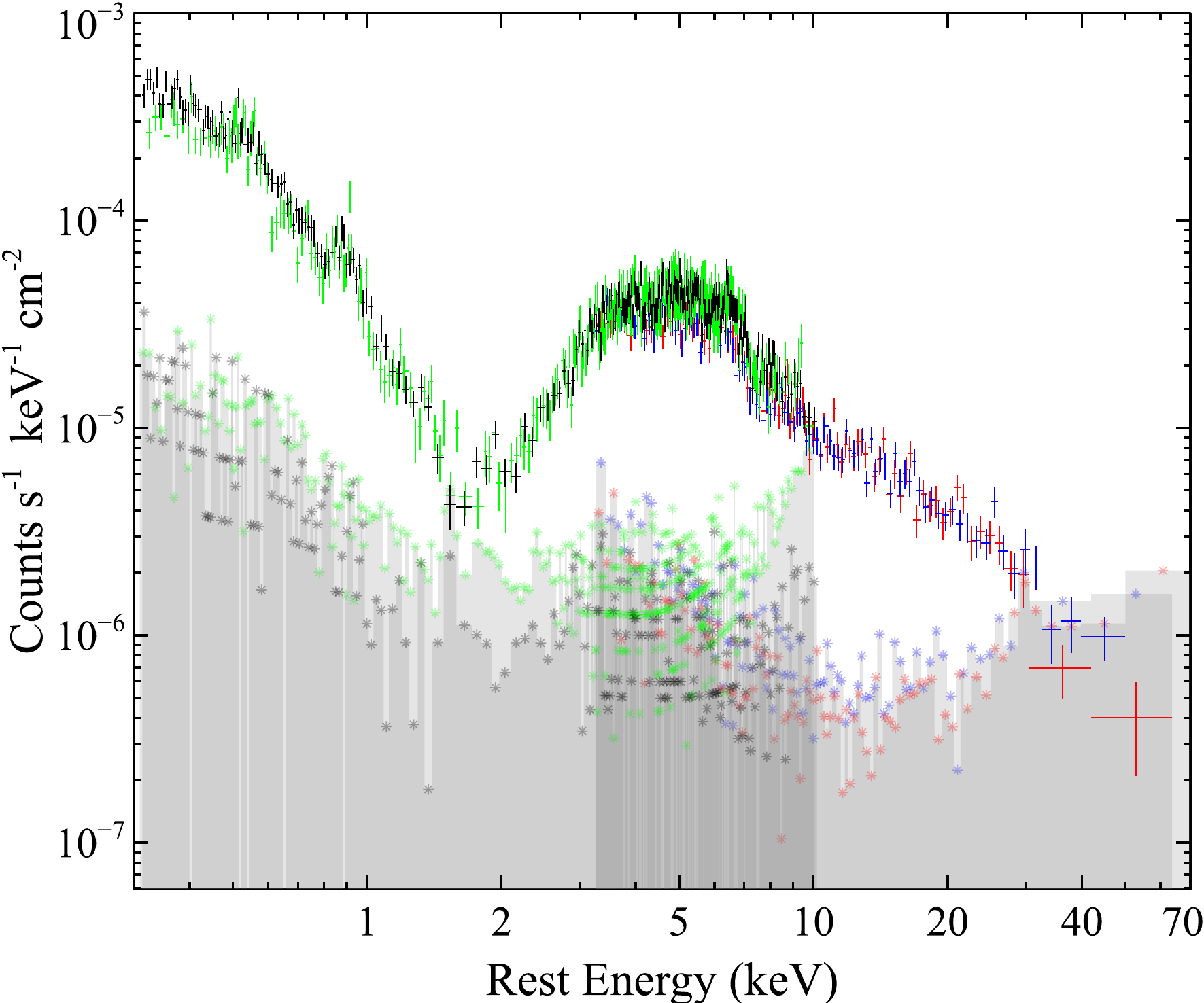}
  %\vspace{-50mm}
  \caption{Time-averaged background-subtracted source and the time-averaged background spectra of the simultaneous \xmmnu observation of \mcg. The net spectra shown here are the pn (black), MOS\,1$+$2 (green), FPMA (red) and FPMB (blue) where the relative background spectra are all plotted with the corresponding colors and clearly dominate above $50\,\kev$. This data have been binned to $3\sigma$ for clarity.}
	\label{fig:mcg03_ldata_bkg}
\end{figure}

\begin{figure*}
  \includegraphics[width=0.70\textwidth]{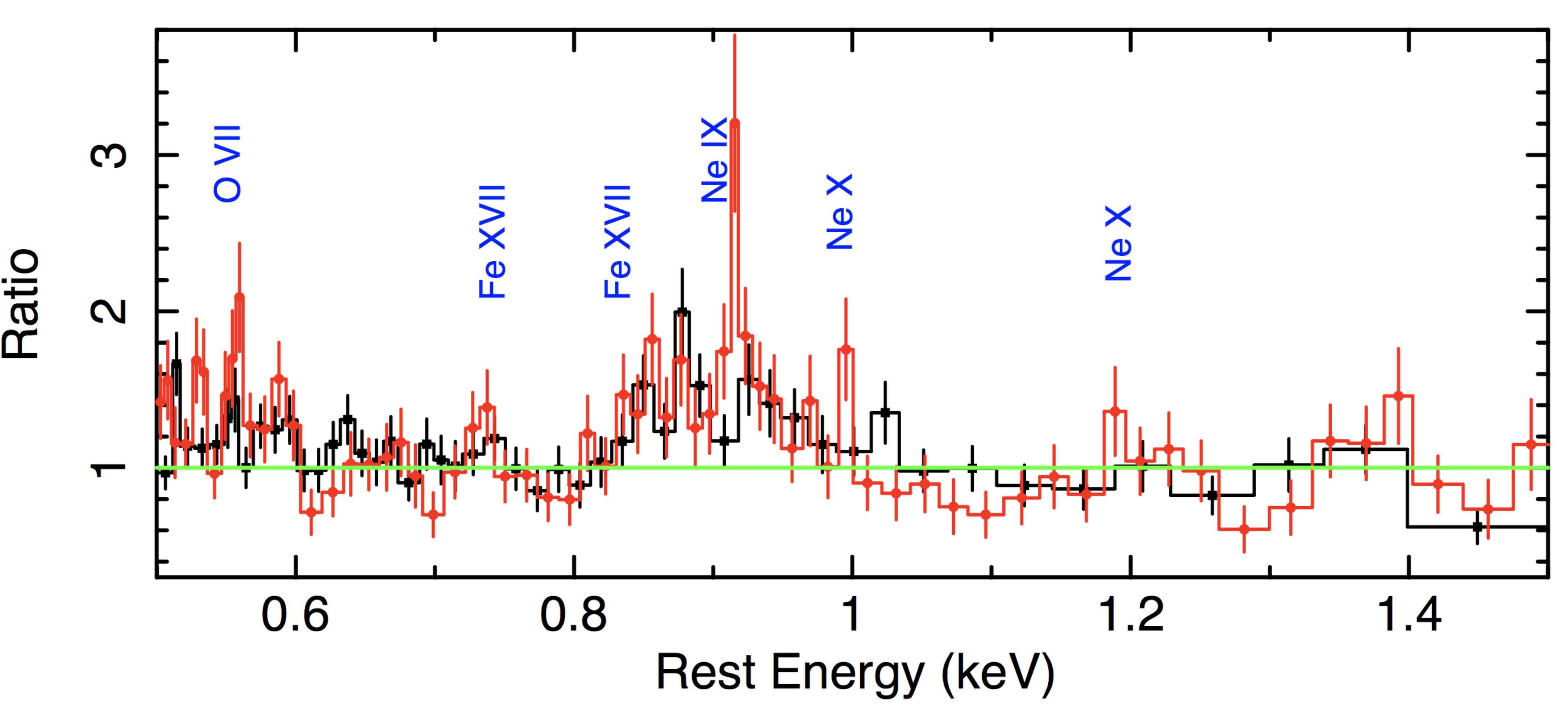}
  %\vspace{-50mm}
  %\includegraphics[width=0.45\textwidth]{figures/mgc03_2slices_eeuf_good}
  \caption{Data/model ratio for the pn (black) and the combined MOS (red) spectra between between $0.5$--$1.5\,\kev$. The adopted model is a simple power-law and Galactic absorption. We note that although \mcg is faint and absorbed in the soft X-ray band, its emission is largely line-dominated.}
\label{fig:mcg3_epic_soft_ra}
\end{figure*}

\begin{figure*}
\centering
\begin{subfigure}[t]{0.6\textwidth}
\centering
\includegraphics[width=\textwidth]{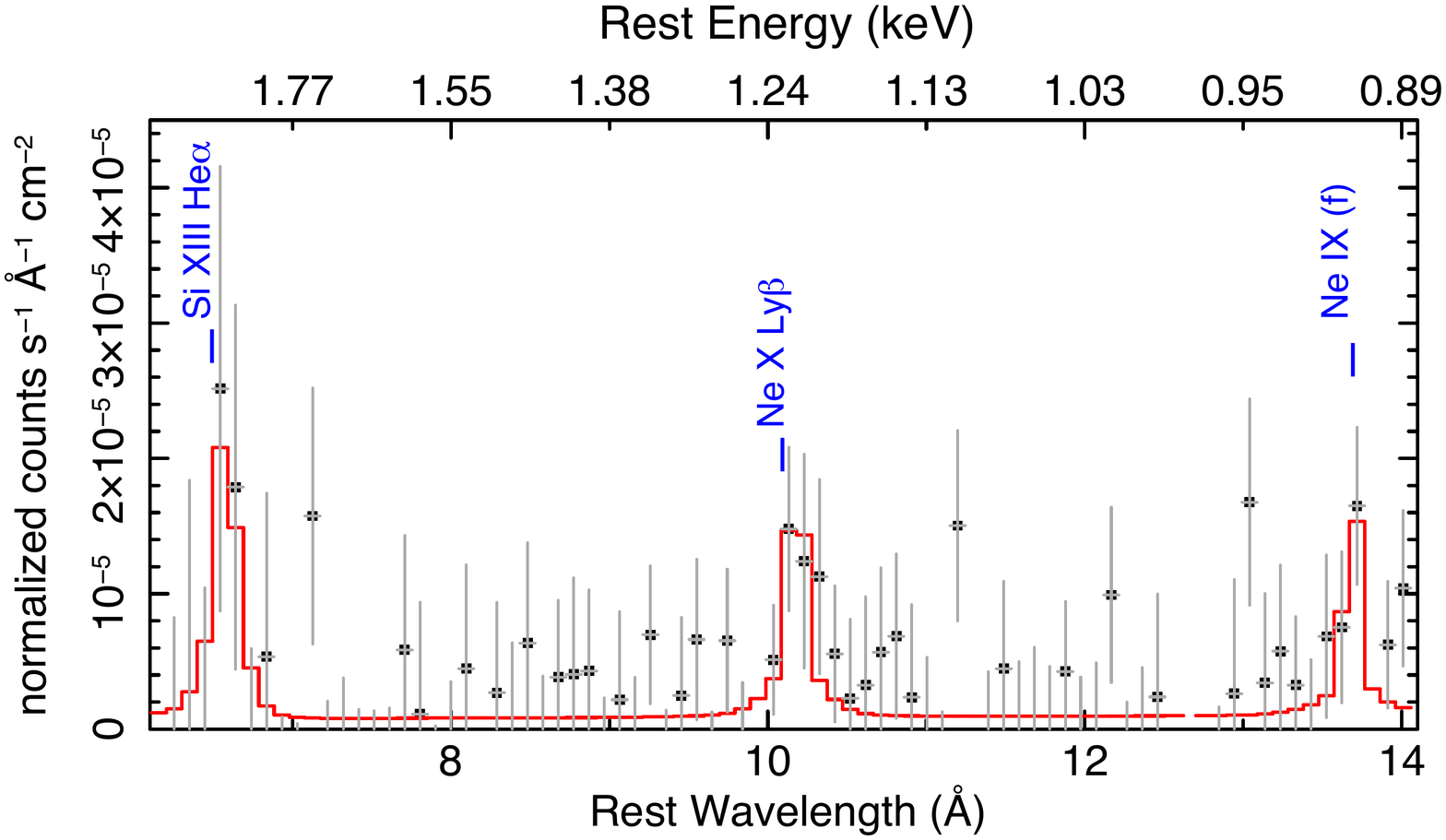} 
%\caption{Generic} \label{fig:timing1}
\end{subfigure}

\vspace{-2.5cm}

\begin{subfigure}[t]{0.6\textwidth}
\centering
\includegraphics[width=\textwidth]{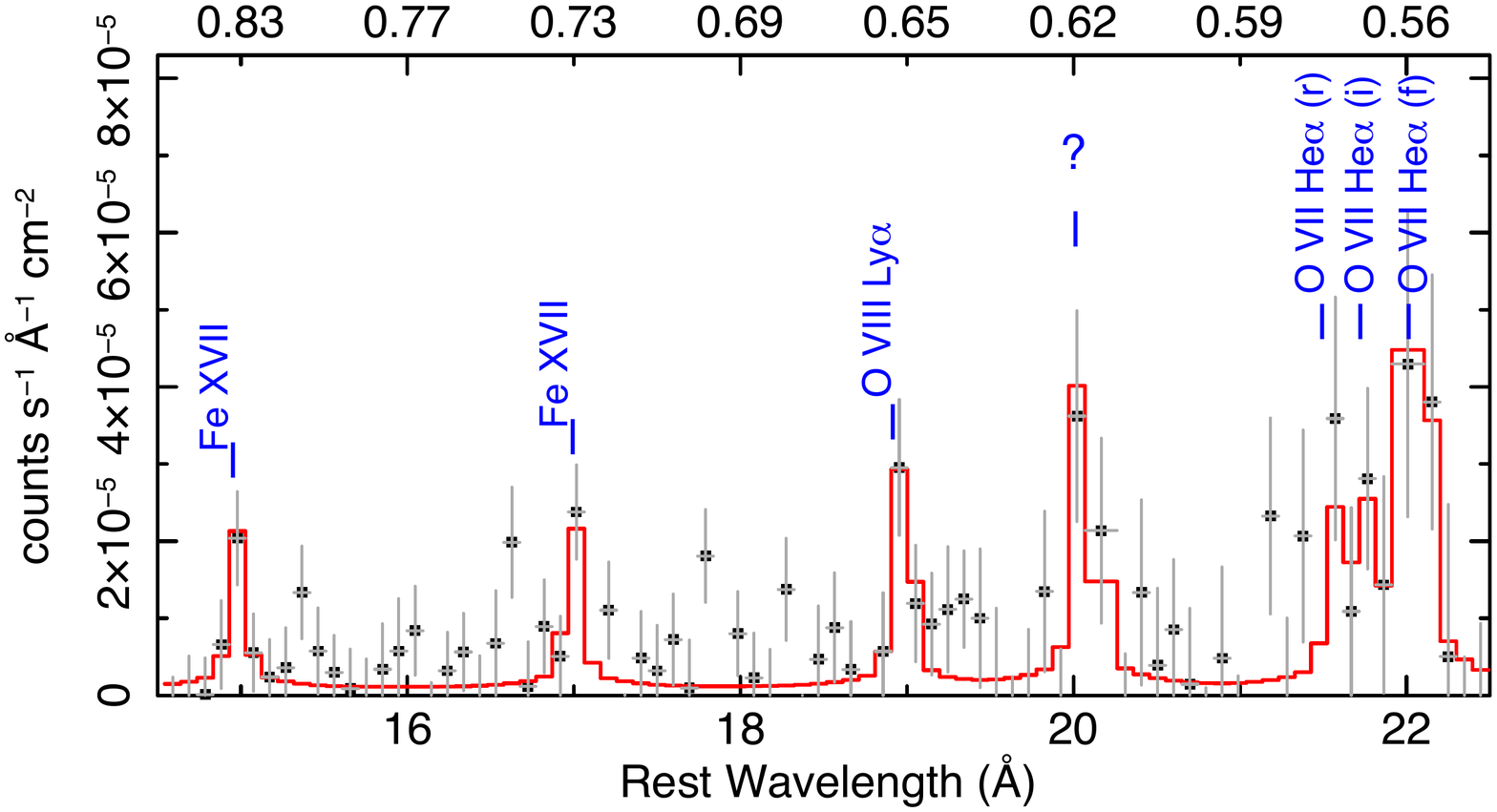} 
%\caption{Competitors} \label{fig:timing2}
\end{subfigure}

\vspace{-2.5cm}

\begin{subfigure}[t]{0.6\textwidth}
\centering
\includegraphics[width=\textwidth]{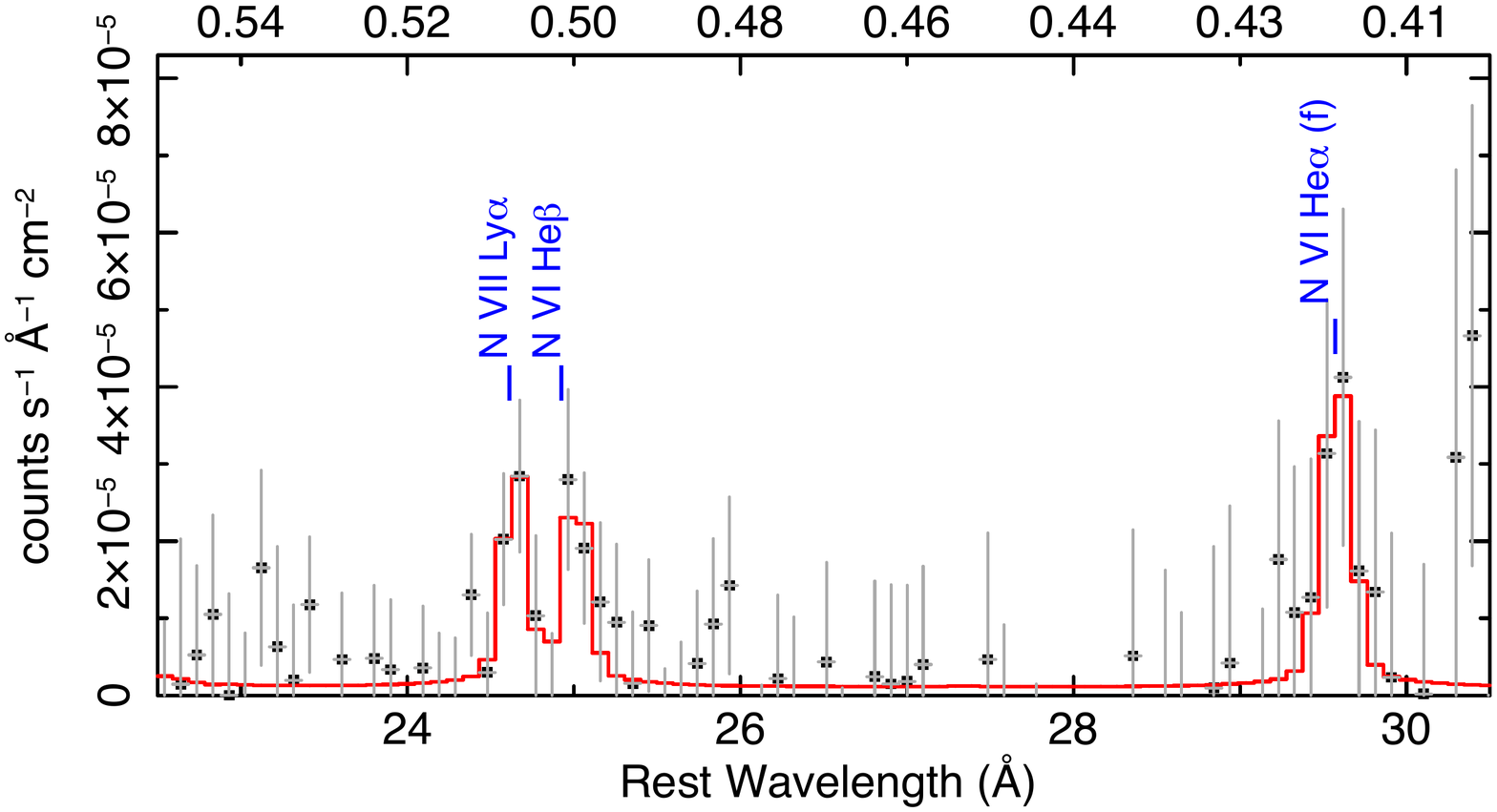} 
%\caption{Price regulation} \label{fig:timing3}
 \end{subfigure}

\vspace{-1cm}

 \caption{Enlarged view of the combined RGS data of \mcg showing the count rate spectrum normalized to the instrumental effective area. The best-fit model is shown by the red line. Despite the low number of counts, it was possible to detect eleven emission lines. Ten out of eleven lines were formally identified with the most likely He- H-like transitions from N, O and Ne. From the plot it is evident that the forbidden transitions are dominating the spectrum over the resonance and intercombination, suggesting the low-density nature of the photoionized distant gas. The emission line significantly detected at rest energy of $E\sim620\,\ev\,(\sim20\,\ang)$ could not be identified. Two additional emission lines were marginally detected at the rest-frame energies of $E=574_{-2}^{+1}\,\ev\,(21.60\ang)$ and $E=1884_{-23}^{+27}\,\ev\,(6.58\ang)$, which are identified to the resonance transitions of O\,\textsc{vii} and Si\,\textsc{xiii} respectively.}
\label{fig:rgs_plot}
\end{figure*}

%\mcg

\section{Observations and Data Reduction}
\label{sec:Observations and Data Reduction}

%\subsection{\xmm}

\xmm observed \mcg simultaneously with \nustar between the 6th--9th of December 2015 with an exposure time of $\sim130\,\ks$ (see Table\,\ref{tab:summary_obs}). The \nustar observation started approximately two hours before the \xmm one and ended $150\,\ks$ afterwards (see Table\,\ref{tab:summary_obs}), with a total duration of $281.8\,\ks$. All the \xmm data were reduced by adopting the Science Analysis System (\sas v16.0.0). The most recent set of calibration files and the EPIC-pn and MOS\,1 $+$ MOS\,2 adopted in this work are the same used in  B18. The same applies for the \nustar data, reduced according to the \nustar Data Analysis Software package (\textsc{nustardas}  v1.6.0). The \xmm RGS data have been reduced by using the standard \sas task \rgsproc, where the high background time intervals were filtered out by applying a threshold of $0.2\,\rm cts\,s^{-1}$ on the background event files. After checking that the RGS\,1 and RGS\,2 spectra were in good agreement, to within the $3\%$ level, we combined them by using the \sas task \rgscombine. The two combined RGS collected a total of $411$ net counts. Due to the low number of net counts, the spectra were binned to $\Delta\lambda=0.1\ang$ which under-samples the RGS FWHM spectra resolution of $\Delta\lambda=0.06$--$0.08\ang$. At this binning, the spectra are characterized by $<20$ counts per channel and hence we adopted C-statistics \citep{Cash79} in the RGS analysis, whereas we adopt $\chi^{2}$ minimization technique for the EPIC and FPMA/B spectral analysis. All the models adopted were fitted to the data by using the standard software package \xspec v12.9.1p \citep{Arnaud96}.

In Fig.\,\ref{fig:mcg03_ldata_bkg} we show the time-averaged pn, MOS\,1$+$\,2 and FPMA/B spectra plotted with the corresponding X-ray background (binned to $3\sigma$ for clarity) between the $0.3$--$70\,\kev$ rest energy band. In the soft band the EPIC data was filtered for high background which severely affected the observation (as reported in B18) while in the hard X-ray band the background dominates above $50\,\kev$. After screening, the background level is only $\sim5$--$10\%$ of the net source rate for the pn and the combined MOS. In the broadband fitting we will confront different physical reprocessing scenarios including the \texttt{MYTorus} model \citep{MurphyYaqoob09}, which is valid at $E>0.6\,\kev$ and hence for the rest of the broadband analysis we concentrate on the $0.6$--$50\,\kev$ band. A Galactic absorption column of the $\nh=2.5\times10^{20}\,\rm cm^{-2}$ \citep{DickeyLockman90}, as from the H\,\textsc{i} $21\,\rm cm$ measurements, was adopted in all the fits and we also assume solar abundance of the main abundant elements throughout the analysis.

\section{Soft X-ray Spectral Analysis}
\label{sec:Soft X-rays Spectral Analysis}

As already shown in B18, both the \suzaku and \xmm soft ($E<2\,\kev$) spectra are rich in emission lines. This is also evident in Fig.\,\ref{fig:mcg3_epic_soft_ra} where we show the $0.5$--$1.5\,\kev$ residuals for the \xmm EPIC data (black and red lines are the pn and MOS spectra, respectively). The adopted model is a simple power-law and Galactic absorption. Since this line-dominated soft spectral range could reveal information on the circum-nuclear gas surrounding the AGN, we analyzed the high resolution spectrum collected with the RGS. In particular, we considered the RGS spectrum between $0.38$--$2\,\kev$ ($6.20$--$32.63\,\ang$) in order to build a template for the soft X-ray emission which we later apply to the broadband and the time-dependent spectral analysis in Section\,\ref{subsec:The broadband xmmnu analysis}.

\begin{table*}
\begin{tabular}{ccccccc}

\hline

E$_{\rm obs}$\,(1)                        &Flux\,(2)                  &$EW$\,(3)               &Transition\,(4)                    &E$_{\rm lab}$\,(5)        &$\Delta C/\Delta\nu$\,(6)\\
\hline
$419_{-1}^{+1}\,[29.59]$                 &$14.7_{-9.9}^{+9.2}$    &$133_{-89}^{+83}$       &N\,\textsc{vi}\,He$\alpha$\,(f)    &$420$                     &$7.0/2$\\
\\
$496_{-2}^{+1}\,[25.00]$                 &$7.0_{-3.8}^{+4.4}$     &$30_{-16}^{+19}$        &N\,\textsc{vi}\,He$\beta$\,(f)     &$495$                     &$11.0/2$\\
\\ 
$503_{-1}^{+1}\,[24.65]$                 &$7.5_{-3.3}^{+3.7}$     &$35_{-15}^{+17}$        &N\,\textsc{vii}\,Ly$\alpha$        &$500.4$                   &$14.3/2$\\
\\
$562_{-1}^{+1}\,[22.06]$                 &$18.6_{-10.0}^{+10.2}$   &$78_{-42}^{+43}$        &O\,\textsc{vii}\,He$\alpha$\,(f)   &$561$                     &$11.1/2$\\
\\
$569_{-2}^{+2}\,[21.79]$                 &$5.0_{-3.7}^{+4.9}$     &$10_{-7}^{+9}$          &O\,\textsc{vii}\,He$\alpha$\,(i)   &$568.7$                   &$4.8/2$\\
\\
%$574_{-2}^{+1}\,[21.60]$                &$<10.73$                &$<23$                   &O\,\textsc{vii}\,He$\alpha$\,(r)   &$573.9$                   &$2.44/2$\\
%\\
$618_{-2}^{+1}\,[20.10]$                 &$12.1_{-6.5}^{+6.6}$    &$229_{-123}^{+124}$     &?                                  &?                         &$10.0/2$\\
\\
$653_{-1}^{+2}\,[18.99]$                 &$6.3_{-2.7}^{+2.9}$     &$169_{-72}^{+78}$       &O\,\textsc{viii}\,Ly$\alpha$       &$653.7$                   &$13.8/2$\\
\\
$729_{-2}^{+1}\,[17.01]$                 &$4.2_{-1.7}^{+2.0}$     &$145_{-59}^{+69}$       &Fe\,\textsc{xvii} 3s\,$\rightarrow$\,2p&$727.1$               &$16.1/2$\\
\\
$828_{-3}^{+2}\,[14.98]$                 &$3.8_{-1.7}^{+1.8}$     &$179_{-81}^{+87}$       &Fe\,\textsc{xvii} 3d\,$\rightarrow$\,2p&$825.8$               &$14.4/2$\\
\\
$906_{-3}^{+2}\,[13.69]$                 &$3.2_{-1.8}^{+1.8}$     &$186_{-104}^{+108}$     &Ne\,\textsc{ix}\,He$\alpha$\,(f)   &$905.1$                   &$8.7/2$\\
\\
$1219_{-14}^{+5}\,[10.17]$               &$4.0_{-2.1}^{+2.2}$     &$460_{-253}^{+260}$     &Ne\,\textsc{x}\,Ly$\beta$          &$1211$                    &$9.2/2$\\
%\\
%$1884_{-23}^{+27}\,[6.58]$              &$4.96_{-4.80}^{+4.78}$     &$>70$                   &Si\,\textsc{xiii}\,He$\alpha$\,(r) &$1865$                 &$2.9/2$\\
%\\
\hline
\end{tabular}
\caption{Summary of the emission lines in the RGS spectra from the best-fit Gaussian model. Note that all the lines are practically unresolved and hence we assume a $\sigma=1\,\rm eV$ during fitting.}
\label{tab:rgs_lines} 
\vspace{-5mm}
\begin{threeparttable}
\begin{tablenotes} 
	\item (1) Measured line energy in the quasar rest frame, in units of eV where the corresponding mean wavelength value in units of $\ang$ is given within the brackets,  
	\item (2) line photon flux, in units of $10^{-6}$\,cm$^{-2}$\,s$^{-1}$,
	\item (3) the equivalent width in the AGN rest frame in units of eV, 
	\item (4) list of the most plausible identification where (f), (i), and (r) refer to forbidden, intercombination, and resonance transitions for He-like species,    
    \item (5) corresponding laboratory energy for the detected lines, in units of eV,	
	\item (6) improvement of the fit in C--statistics upon adding the line to the model.\end{tablenotes}
\end{threeparttable}
\end{table*}

\subsection{RGS analysis}
\label{subsec:RGS analysis}

As a first step, the RGS spectrum was fitted with a simple power-law (\texttt{zpowerlaw}) and a Galactic absorption model (\texttt{Tbabs} with cross-section and ISM abundances of \citealt{Wilms00}). This simple continuum model returned a photon-index of $\Gamma\sim2.4$ which is broadly consistent with that found in the broadband analysis in Section\,\ref{subsec:The broadband xmmnu analysis}. Since the residuals in the spectrum suggests the presence of multiple unresolved narrow emission lines, we modelled them with Gaussian components (\texttt{zgauss}) in \xspec, with the line-width fixed at $\sigma=1\,\rm eV$. We note that, the inclusion of a line requires to provide an improvement of $\Delta C \geq 4.6$ for $2$ free parameters (i.e., line energy and the normalization) corresponding to a $\geq 90\%$ confidence level. Following this criterion we formally detected eleven emission lines, ten of which have a secure identification mostly associated, in their rest frame, with He- and H-like of abundant elements such as N, O and Ne. There is an additional line at $618_{-2}^{+1}\,\ev$ (or $20.10\ang$) which does not correspond to a known transition. Two additional emission lines where marginally detected ($\dC<4.6/2$) at the expected rest-frame energies of $E=574_{-2}^{+1}\,\ev\,(21.60\ang)$ and $E=1884_{-23}^{+27}\,\ev\,(6.58\ang)$ which can be associated to the resonance transition of O\,\textsc{vii} and Si\,\textsc{xiii} respectively.

The overall fit statistic considerably improved upon the addition of the Gaussian line profiles decreasing from $\C=407/289$, without any line emissions (null hypothesis probability $2.6\times10^{-5}$) to a more statistically acceptable $\C=298/263$ (null hypothesis probability $9.8\times10^{-2}$) with the included lines. Fig.\,\ref{fig:rgs_plot} shows the rest frame $6$--$30\ang$ RGS spectrum with the best-fitting model (red lines) overlaid on. The emission line fluxes, equivalent widths ($EW$) and identifications are listed in Table\,\ref{tab:rgs_lines}. The centroid energy of the O\,\textsc{vii} line at the rest energy of $E=562\pm1\,\ev$\,($22.06\ang$) suggests an identification with the forbidden transition, detected at $3\sigma$ confidence level. The intercombination transition is also detected at the expected energy of $E=569\pm2\,\ev\,(21.79\ang)$ but with a lower significance at the $\sim2\sigma$ level. %On the other hand, in this O\,\textsc{vii} triplet, the recombination emission is tentatively detected ($1\sigma$) at $E=574_{-2}^{+1}\,\ev\,(21.60\ang)$. 

As for the other He-like complexes such as the N\,\textsc{vi} and Ne\,\textsc{ix} triplets, we only observed the forbidden component at the expected energy of $E=419\pm1\,\ev\,(29.59\ang)$ and $E=906_{-3}^{+2}\,\ev\,(13.69\ang)$. Although the RGS spectrum has a low-number of counts, we can observe that the soft X-ray emission lines are largely dominated by forbidden transitions rather than resonance and/or intercombination emission (see Fig.\,\ref{fig:rgs_plot}).

This result suggests that such emission line mainly originates from a distant low-density photoionized plasma as seen in many Seyfert\,2s \citep[e.g.,][]{Sako00,Kinkhabwala02,Kallman14}. In particular, a strong emission line well match with the O\,\textsc{viii}\,Ly$\alpha$ emission at the expected rest frame energy of $E=653_{-1}^{+2}\,\ev\,(18.99\ang)$. The line width estimate ($\sigma<3.7\,\ev$) corresponds to an upper limit on the velocity of $\sigma_{\rm v}\lesssim1700\,\kms\,(FWHM\lesssim4000\,\kms)$ and hence broadly consistent with emitting gas possibly located in the BLR/NLR.

A possible diagnostic for the density of the emitting gas can be carried out by quantifying the ratio between the intensity of the forbidden and intercombination components detected in He-like triplets \citep{PorquetDubau00}. However in this work, the only possible diagnostic can be carried out on the O\,\textsc{vii} He$\alpha$ lines as this is the only triplet detected. The resulting emission line ratio is $R=z/(x+y)=3.7_{-0.8}^{+1.6}$ (where $z$ and $x+y$ are the forbidden and intercombination transition respectively). This ratio places an upper limit on the electron density of the order of $n_{\rm e} < \rm 2\times10^9 cm^{-3}$ (see Fig.8 \citealt{PorquetDubau00}). Nevertheless, the spectral resolution in the RGS does not allow to put accurate constraints on the widths and location of these lines. The above estimates suggests that the lines could be, in principle, broad and emitted from a gas with a density lower than $\times10^9\,\rm cm^{-3}$ and hence located at $R_{\rm gas}\gtrsim$\,BLR.

Furthermore, all the detected soft emission lines are consistent to their expected rest frame energies, except N\,\textsc{vii}\,Ly$\alpha$ line which lies slightly above it (see Table\,\ref{tab:rgs_lines}). The possible emission line detected at $E=729_{-2}^{+1}\,\ev\,(17.01\ang)$, has an uncertain identification but it could be associated with an L-shell transition from iron (e.g., Fe\,\textsc{xvii} 3s\,$\rightarrow$\,2p). A contribution from the radiative recombination continuum (RRC) of O\,\textsc{vii} is also plausible. Another strong iron\,L emission is identified at rest energy of $E=828_{-3}^{+2}\,\ev\,(14.97\ang)$, which lies very close to the expected rest-frame energy of Fe\,\textsc{xvii} 3d\,$\rightarrow$\,2p transition. The iron\,L transitions are associated with collisionally ionized plasma, likely within the star forming regions in the host galaxy.

Having reached a good parameterization of the narrow emissions with the Gaussian profiles, we subsequently replaced the Gaussian emission line fit with a publicly available \xstar photoionization emission grid\footnote{ftp://legacy.gsfc.nasa.gov/software/plasma$\textunderscore$codes/xstar/xspectables/} \citep{BautistaKallman01}. Despite the low-number of counts in the RGS, the inclusion of the \xstar grid can provide a first order estimate on the physical condition of the gas such as ionization, column and electron density. As the emission profiles appear to be narrow, we chose a suitable grid with velocity broadening of $\vturb=100\,\kms$. Moreover, we find that in order to model it with \xstar, it is preferable to adopt a larger binning (i.e., $\Delta\lambda=0.2\ang$) to the spectrum.

From the photoionization modelling, the \xstar emission flux (or normalization) is expressed as: 

\begin{equation}
\kappa_{\rm xstar}=f_{\rm cov}\dfrac{L_{38}}{D_{\rm kpc}^{2}}
\end{equation}

\noindent where $L_{38}$ is the ionizing luminosity ($\lion$) in units of $10^{38}\,\ergs$, $D_{\rm kpc}$ is the luminosity distance of the ionizing source (AGN) in kiloparsecs, and $\fcov=\Omega/4\upi$ is the covering fraction of the gas. Thus by keeping the $\nh$ fixed to the default value given in the \xstar table of $\lognh=21.5$, which is more typical of a diffused warm emission component found in Seyfert\,1 and Seyfert\,2s galaxies \citep[e.g.,][]{Kinkhabwala02,Blustin05}, we get a normalization of $\kappa_{\rm xstar}=1.2_{-0.4}^{+0.4}\times10^{-6}$. By assuming a bolometric luminosity in \mcg of $\lbol\sim3\times10^{45}\,\ergs$ (B18) and that $\lion\sim\lbol/3\sim10^{45}\ergs$, at a luminosity distance of $D_{\rm L}=138.8\,\rm Mpc$, we obtain a corresponding gas covering fraction of the order of $\sim0.2\%$. This suggests that the distant photoionized gas is covering a small fraction of the sky and possibly inhomogeneous, however this result is strongly dependent on a given column density. If we assume a $10\%$ covering factor, which is broadly expected in the NLR \citep[e.g.,][]{NetzerLaor93}, it would require a normalization of $\kappa_{\rm xstar}=5.2\times10^{-5}$ corresponding to a column density of $\lognh=19.9_{-0.2}^{+0.1}$.

\begin{figure*}
\includegraphics[scale=0.60]{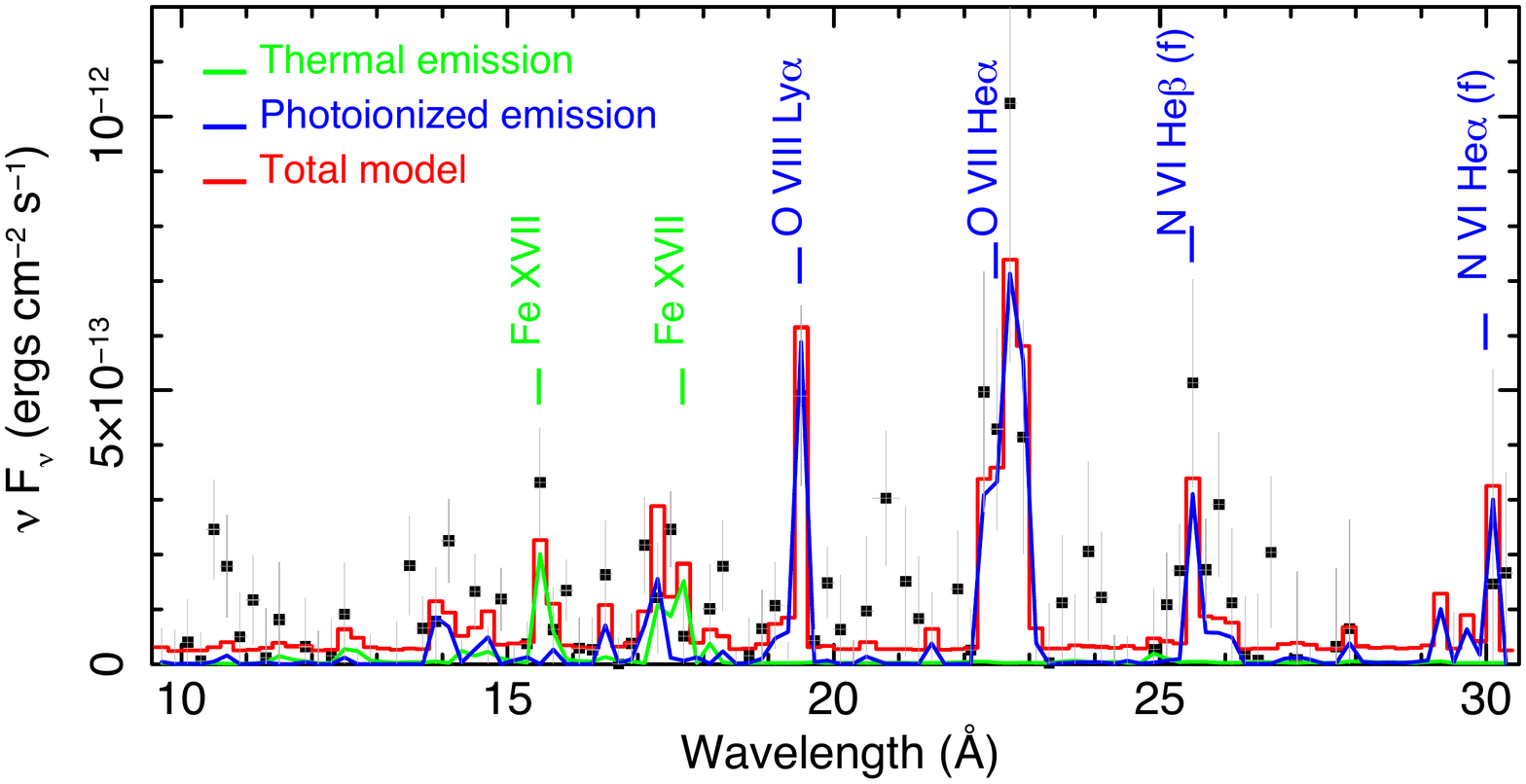}
  \vspace{-15mm}
\caption{Best-fit model (red) overlaid on the RGS spectrum with group $\Delta\lambda=0.2\ang$ for extra clarity. The corresponding contribution of both thermal (green) and photoionized (blue) components are shown, where the former is responsible for the collisionally ionized Fe\,L emission between $15$--$17\ang$ and the latter mainly responsible to all the forbidden transition emission e.g., N\textsc{vi} and O\textsc{vii} He$\alpha$. The RGS spectrum is clearly dominated by emission from photoionized elements from distant gas possibly located in the NLR.}
\label{fig:rgs_xstar_plot}
\end{figure*}

The ionization state of the plasma was measured to be $\logxi=1.18_{-0.27}^{+0.22}$ and the addition of the photoionized gas emission component results in a considerable improvement to the overall fit to $\dC=34.2/2$ which is at $>99.99\%$ confidence level. This is somewhat expected considering that the soft X-ray emission is dominated by forbidden transitions. The inclusion of a thermal component via the \xspec \citep[\texttt{mekal}; ][]{Mewe85} model, led the a modest improvement in the fit statistic of $\dC=8.5/2\,(2.5\sigma)$. Here we find a poorly constrained temperature of $kT=0.35_{-0.15}^{+0.25}\,\kev$ and a normalization of $1.5_{-0.9}^{+0.4}\times10^{-5}\,\frac{10^{-14}}{4\upi[D_{A}(1+z)]^2}  \int n_{\rm e} n_{\rm H}dV$, where $n_{\rm e}$ and  $n_{\rm H}$ are the electron and hydrogen densities (measured in $\rm cm^{-3}$) respectively and $D_{A}$ is the angular diameter of the source in $\rm cm$. Lastly, we kept the photon-index of the soft power-law component fixed at the best value, found in B18 and later in the broadband analysis of $\Gamma=2.2$. Note that with such a complex model, the normalization of the scattered power-law component is no longer well constrained. In Fig.\,\ref{fig:rgs_xstar_plot} we show the best-fit model (red) to the data with the corresponding contribution of both thermal (green) and photoionized (blue) components. The latter clearly dominate across the RGS spectrum.

\begin{figure*}
\includegraphics[scale=0.50]{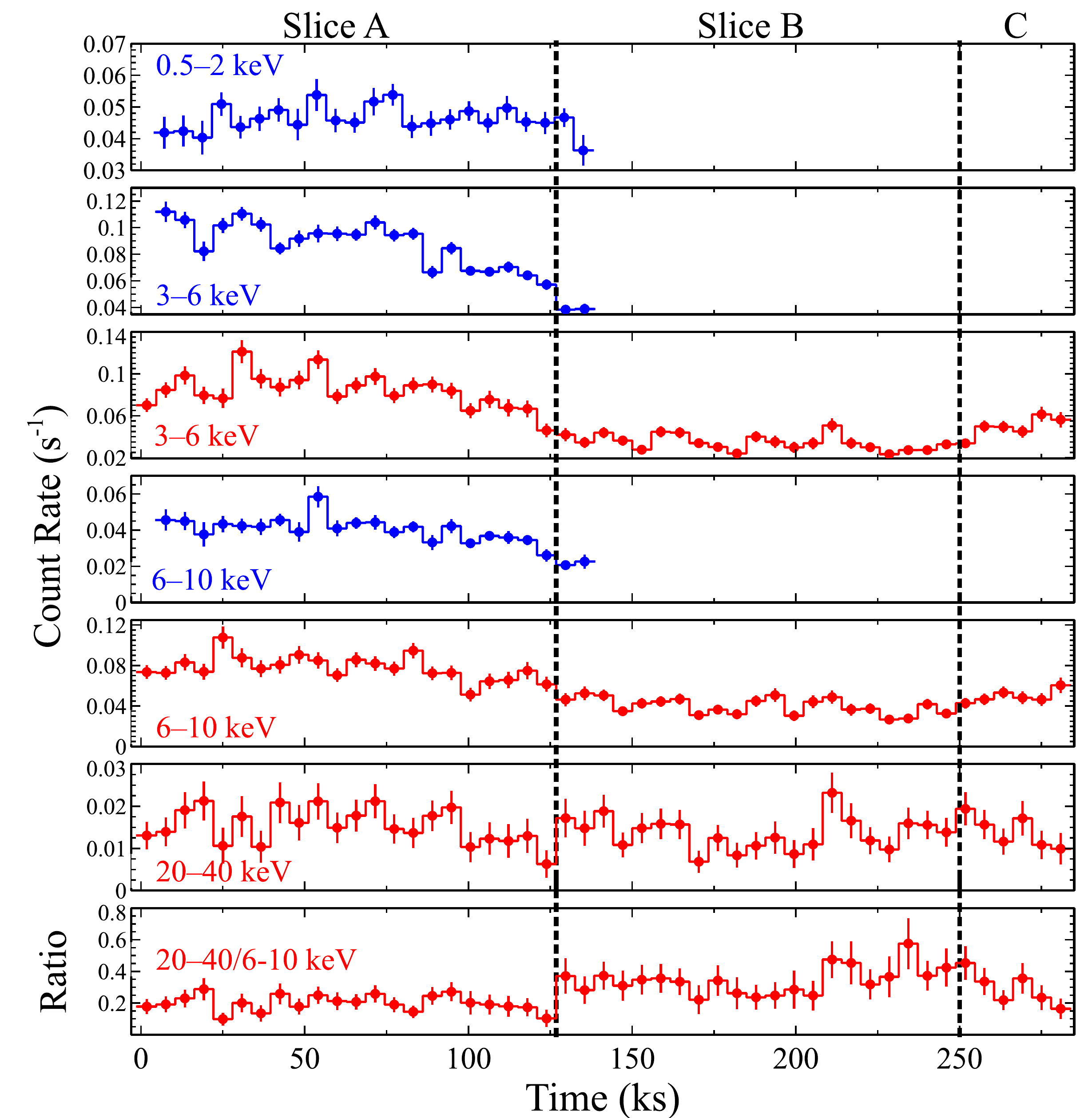}
%  \vspace{-15mm}
\caption{Lightcurve in different energy bands (i.e., $0.5$--$2\,\kev$ $3$--$6\,\kev$ $6$--$10\,\kev$ and $20$--$40\,\kev$) corresponding to EPIC-pn (blue) and FPMA$+$B (red) both binned at 5814 seconds (one \nustar orbit) respectively. The hardness ratio between $20$--$40\,\kev$ and $6$--$10\,\kev$ corresponding to the FPMA$+$B is plotted on the bottom panel. \mcg was observed simultaneously with \xmmnu, the duration of the latter extended $150\,\ks$ beyond \xmm to $281.8\,\ks$. According to the behavior of the lightcurves, the \xmm and \nustar spectra were sliced in three segments named slice\,A--C which are visually marked on the plot by the vertical black dotted lines. Thus slice\,A and slice\,B are characterized by the \xmm (pn and MOS\,1$+$2) and \nustar (FPMA$+$B) spectra separated at $0$--$125\,\ks$ and $125$--$250\,\ks$ respectively. However, in slice\,B we are only left with less than $10\,\ks$ of \xmm data which implies that the \xmm contribution will be only from a short EPIC-pn spectrum. Furthermore, slice\,C is entirely consisting of \nustar data between $250$--$280\,\ks$.}
	\label{fig:mgc03_nustar_xmm_lc}
\end{figure*}
%\pagebreak

\section{Broadband \xmmnu analysis}
\label{subsec:The broadband xmmnu analysis}

\subsection{Description of the \xmmnu lightcurves}
\label{subsec:Description of the xmmnu lightcurves}

In Fig.\,\ref{fig:mgc03_nustar_xmm_lc} we show the \xmm (blue) and \nustar (red) lightcurves in different energy bands, as well as the hardness ratio between the $6$--$10\,\kev$ and $20$--$40\,\kev$ bands observed by \nustar. The lightcurve in the $0.5$--$2\,\kev$ band shows no variability while the flux in the $3$--$6\,\kev$ band in particular drops strongly at around 125\,ks, from $\sim11\,\rm cts\,s^{-1}$ to $\sim0.04\,\rm cts\,s^{-1}$, after the start of the \nustar observations. In comparison the $20$--$40\,\kev$ hard X-ray band shows little variation throughout the observations. Indeed the ratio between the $20$--$40\,\kev$ and $3$--$6\,\kev$ bands shows a clear increase in hardness at this point, which B18 earlier interpreted as an obscuration event which occurs within an approximate timescale of one day. Then at the end of the observation at 250\,ks, the source flux gradually starts to recover in the remaining 30\,ks of the \nustar observation.

Accordingly to this behavior and following B18, the observations were split into three time intervals; from $0$--$125\,\ks$\,(slice\,A), $125$--$250\,\ks$\,(slice\,B) and from $250$--$280\,\ks$\,(slice\,C), as marked by vertical dotted lines in Fig.\,\ref{fig:mgc03_nustar_xmm_lc}. Slice\,A has a high count rate and is largely coincident with the whole of the \xmm observation. On the other hand, in Slice\,B captures the increase in hardness of the source, where in particular the $3$--$6\,\kev$ band count rate drops compared to the $20$--$40\,\kev$ lightcurve. Note that slice\,B includes less than $10\,\ks$ of \xmm data, which is only included to constrain the soft band spectrum, below $3\,\kev$. In slice\,C the hardness ratio drops again whereas the source brightness appears to recover. Slice\,C only consists of \nustar data between $250$--$280\,\ks$ and the spectral properties are intermediate between slice\,A and B. As the quality of spectrum for Slice\,C is low (with only $\sim30\,\ks$ of net exposure for both \nustar FPM detectors combined), we do not include this in our subsequent quantitative analysis and thereafter concentrate on Slice\,A and Slice\,B. Nonetheless, the overall variability provides important physical constraints, which as suggested by B18, indicates that the AGN went through an absorption event in order to account for the rapid change in hardness ratio.

\subsection{Broadband analysis of slice\,A}
\label{subsec:Analysis of slice A}

For the initial broadband analysis we consider slice\,A rather than the time-averaged spectra as the former includes the only time-interval where both telescopes are effectively observing simultaneously (see Fig.\,\ref{fig:mgc03_nustar_xmm_lc}). Since we obtained a reasonable fit with \xstar and \mekal in the RGS (see \ref{subsec:RGS analysis}), we adopt the same best-fit values in the EPIC broadband spectra. This was achieved by fixing the ionization parameter and by letting the corresponding normalizations readjust accordingly including the temperature of the \mekal component.

In this work we adopt two main models. Model\,$\mathcal{A}$ represents a more classical model where the primary continuum is modelled as power-law component transmitted through a neutral absorber. The Compton reflected component is produced by a distant neutral reflector which is geometrically approximated as a slab of neutral material with \texttt{pexmon} \citep{Nandra07}, whilst a distant scattered continuum component is also included throughout. Model\,$\mathcal{B}$ assumes a geometrically toroidal reproccesor modelled with \texttt{MYTorus} \citep{MurphyYaqoob09}, assuming a standard (coupled) configuration, which takes into account the physical properties of the absorbing medium, the Compton-down scattering effect and it includes self-calculated reflected components (continuum plus Fe\,K emission lines). This model also assumes a fixed geometry of the toroidal X-ray reprocessor, a single value for the covering factor of the torus (corresponding to a half--opening angle of 60$^\circ$) and a uniform composition of the torus itself. Model\,$\mathcal{B}$ will be described in Section\,\ref{subsubsec:xmmnu modeling with MYTorus}. In the following we adopt model\,$\mathcal{A}$, defined as:

\begin{equation*}
\begin{split}
\rm Model\,\mathcal{A}=\texttt{Tbabs}\times[\texttt{zpowerlw}_{\rm scatt}+\texttt{xstar}_{\rm em}+\texttt{mekal}\\+ \,\texttt{pexmon}+(\texttt{xstar}_{\rm FeK,1}\times\texttt{xstar}_{\rm FeK,2})\\\times\texttt{zpowerlw}_{\rm intr}\times\texttt{zphabs}],
\end{split}
\end{equation*}

\noindent where \texttt{Tbabs} accounts for the Galactic absorption, the scattered power-law component is parameterized with \texttt{zpowerlw}$_{\rm scatt}$. \texttt{zpowerlw}$_{\rm intr}$ models the primary continuum which is absorbed by fully covering neutral material (\texttt{zphabs}) with a column density of $\lognh=23.38\pm0.02$. The Compton-reflection component is modelled with \texttt{pexmon} \citep{Nandra07} which includes the power-law continuum reflected from distant neutral material and the emission from \feka, \fekb, Ni\,K$\alpha$ and \fe Compton shoulder. The photon-indexes of the reflected and primary continuum components are assumed to be the same, the inclination angle was fixed as $60^{\circ}$ ($0^{\circ}$ corresponds to face-on), and the cutoff energy was $300\,\kev$. The normalization was allowed to vary and the scaling reflection factor parameter was fixed at $R=\Omega/2\upi=1$.

B18 detected in the \xmmnu spectra two strong absorption features at rest frame energies of $7.4\pm0.1\,\kev$ and $10.2\pm0.1\,\kev$ associated to blueshifted $1s\,\rightarrow\,2p$ transition of \fexxv and \fexxvi. These absorption profiles likely correspond to two zones of highly ionized outflowing absorbers with velocities of $\vw\sim -0.1c$ and $\vw\sim -0.3c$. Since the parameterization with the Gaussian absorption profiles has been already explored in B18, here we simply model these features with two multiplicative grids of photoionized absorbers generated with \xstar photoionization code \citep[v2.21bn13,][]{Kallman04}, defined as \texttt{xstar}$_{\rm FeK,1}$ and \texttt{xstar}$_{\rm FeK,2}$ respectively. Since these lines are broad, we adopt for all the subsequent fits the original grid used for \pds \citep{Nardini15,Matzeu16} with a high velocity broadening of $\vturb=10000\,\kms$. In generating this grid, it was also assumed an intrinsically steep ionizing continuum of $\Gamma=2.4$, which is consistent with what found in \mcg. We find that the addition of two absorption zones with outflow velocities of $\vwc=-0.09\pm0.02$ and $\vwc=-0.35\pm0.02$ and column density of $\lognh=23.2\pm0.1$ and $\lognh=24.0_{-0.2}^{+0.3}$ improved the fit by $\dchis=28.8/3$ ($4.7\sigma$) and $\dchis=19.9/3$ ($3.7\sigma$) for zone\,1 and zone\,2 respectively. The slower zone prefers a lower ionization ($\logxi=5.4\pm0.2$) than the faster zone ($\logxi=6.2_{-1.0}^{+0.4}$). For the latter zone, the ionization and column density of the ionized absorber are highly degenerate which explains the large error. Here we note that the \xmmnu spectra are dominated by the transmitted primary absorbed component above $\sim2\,\kev$. The intensities of the reflected and primary component are $\rm norm_{\rm pexmon}=1.3\pm0.5\times10^{-3}\,\knorm$ and $\rm norm_{\rm zpow}=2.9_{-0.3}^{+0.4}\times10^{-3}\,\knorm$ respectively which corresponds to a reflection fraction $R=0.5\pm0.1$. Overall model\,$\mathcal{A}$ provided a good fit to slice\,A of $\chis=714/660$. The details of the best-fit model are listed in Table\,\ref{tab:broadband_models_slice1}.

In terms of the soft excess, the soft photoionized emission component is modelled with the same \xstar emission grid (\texttt{xstar}$_{\rm em}$) and the corresponding $\nh$ and $\xi$ values used in the RGS. Its normalization is readjusted to a slightly higher measurement (but consistent within the errors found in the RGS) of $\kappa_{\rm xstar}=1.6_{-0.5}^{+0.4}\times10^{-6}$. If we let the ionization parameter to be free to vary, the resultant fit would lead to a considerable decrease of $\xi$ from $\logxi\sim1.2$ to $  \logxi\sim0.6$. However if we assume this value in the RGS model it would would practically account only for the steep continuum and not model than the individual emission lines. The other contribution to the soft excess is associated with starburst emission which is modelled with a thermal component emission (\texttt{mekal}) with a temperature of  $kT=0.79_{-0.05}^{+0.05}\,\kev$. We find that after including both the thermal and the photoionized components, the photo-index of the scattered continuum is still steep at $\Gamma=3.3_{-0.8}^{+0.9}$. Note that such steep value cannot be decreased even by adding a second \mekal component. This might indicate that the scattered soft X-ray power-law is steeper than what observed in the hard X-rays. This could be caused by the presence of an intrinsic soft excess as seen in many type\,1 AGN \citep[e.g.,][]{Singh85,TurnerPounds88,Nardini11}.

\subsubsection{\xmmnu modelling with MYTorus}
\label{subsubsec:xmmnu modeling with MYTorus}

In the above fitting, the \texttt{pexmon} model assumes geometrically a simple slab reflector. In the following we replace both the \pexmon and the simple neutral absorber (\texttt{zphabs}) with the \myt model. It is essentially composed by three tables, developed for \xspec, of reprocessed spectra \citep[for more details see][]{MurphyYaqoob09} assuming a primary power-law input spectrum that interacts with a reprocessor with toroidal geometry. We therefore analyze slice\,A with model\,$\mathcal{B}$ which is constructed mathematically as follows:

\begin{equation*}
\begin{split}
\rm Model\,\mathcal{B}=\texttt{Tbabs}\times[\texttt{zpowerlw}_{\rm scatt}+\texttt{xstar}_{\rm em}+\texttt{mekal}\\+ \,A_{\rm S}\times\texttt{MYTorusS}+\texttt{MYTorusZ}\\\times(\texttt{xstar}_{\rm FeK,1}\times\texttt{xstar}_{\rm FeK,2})\times\texttt{zpowerlw}_{\rm intr}\\+A_{\rm L}\times\texttt{gsmooth}\times\texttt{MYTorusL}],
\end{split}
\end{equation*}
\noindent where \texttt{MYTorusS} and \texttt{MYTorusL} are publicly available emission grids that reproduce respectively to the reflected continuum and the \feka, \fekb emission line spectrum. \texttt{MYTorusZ} is a multiplicative table corresponding to the zeroth-order transmitted continuum, containing pre-calculated transmission factors that affect the incident continuum due to photoelectric absorption. The \texttt{gsmooth} component is a convolution model, available on \xspec, which takes into account the broadening of the \fe emission lines. The soft excess and the \iron absorption features have been modelled as in model\,$\mathcal{A}$ returning consistent values as listed in Table\,\ref{tab:broadband_models_slice1}.

\begin{figure}
\includegraphics[scale=0.5]{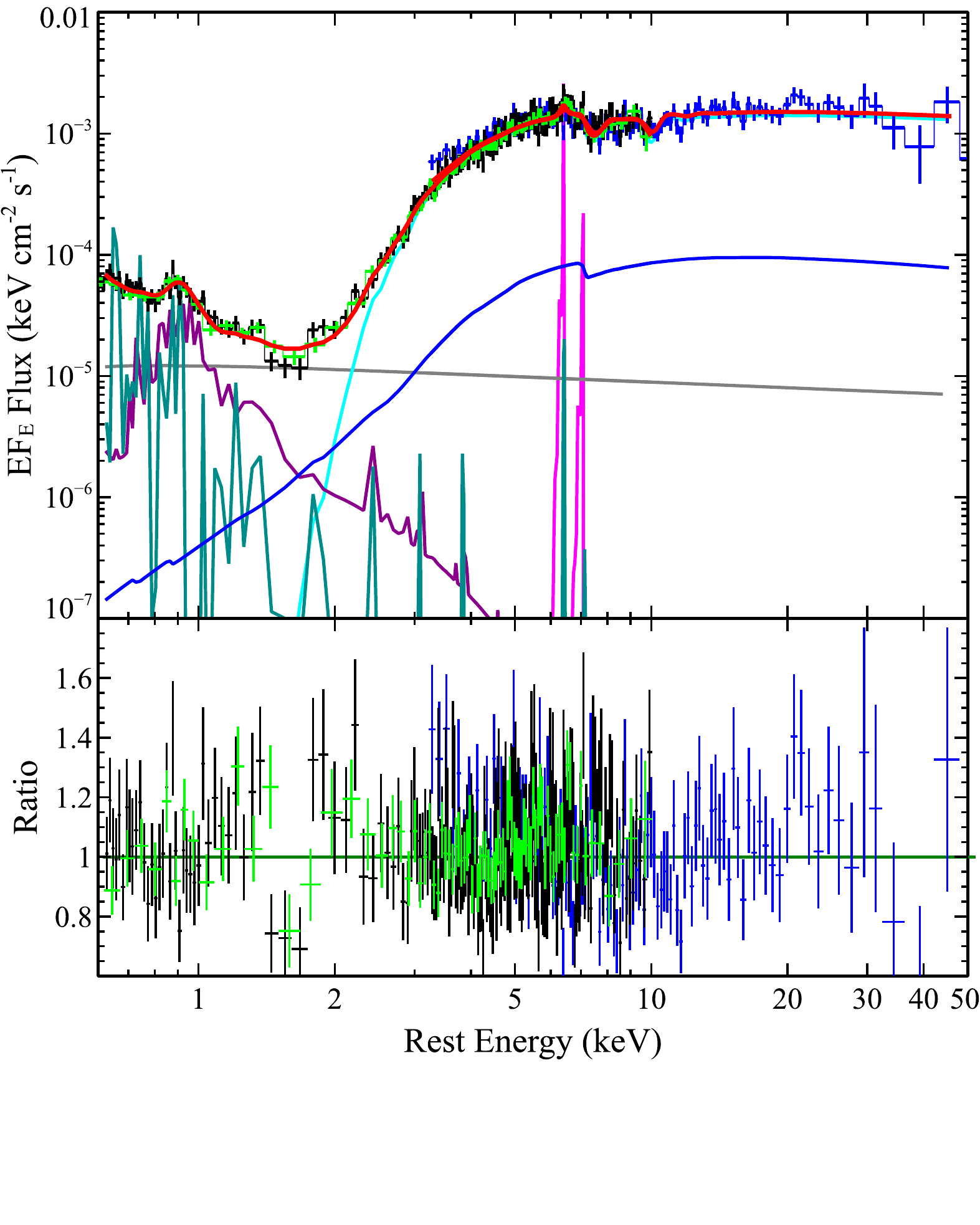}
  \vspace{-15mm}
\caption{Top panel: The simultaneous \xmmnu data of \mcg corresponding to slice\,A. The fluxed spectra was unfolded against a simple power-law with $\Gamma=2$ with the best-fit model\,$\mathcal{B}$ superimposed shown in red. The pn, MOS\,1$+$2 and FPMA$+$B spectra are shown in black, green and blue respectively where the latter have been only co-added for plotting purposes. The contribution of model\,$\mathcal{B}$ are: distant scattered component (gray), photoionized emission (dark cyan), hot thermal emission (dark magenta), whereas the corresponding \texttt{MYTorus} model contributions are the (zeroth-order) transmitted absorbed primary component (cyan), the reflected component (blue) and the \feka and \fekb fluorescent line emission lines (magenta). Bottom panel: The corresponding data/model ratio which shows some residual left $>10\,\kev$.}
\label{fig:mcg03_MYTdec_sliceA}
\end{figure}

Here we constructed \myt assuming the standard `coupled' configuration \citep{Yaqoob12} which assumes that the angle at which the \los directly intercepts the torus is the same (i.e., coupled) as the scattered one. Here the relative reflected and fluorescence line emitting normalizations are set to be equal to the primary power-law normalization i.e., $\rm norm_{\rm S}=norm_{\rm L}=norm_{\rm intr\,zpowl}$, while all the constant factors are set to unity i.e., $\rm A_{\rm S}=A_{\rm L}=1$. The \los inclination angle in respect to the polar axis was fixed at its best-fit value of $\theta_{\rm obs}=70.0^{\circ}$, consistent with the type\,2 classification. The overall column density measured in this (coupled) configuration indicates that we are viewing the source through Compton thin material i.e., $\lognh=23.31\pm0.01$, which might suggest that the upper regions of the torus are indeed less dense than the more central one. Overall in model\,$\mathcal{B}$ the transmitted component also dominates over the reflected component above $2\,\kev$. The photo-index was found to be $\Gamma=2.14\pm0.05$ and the normalization of the primary is $2.7\pm0.3\times10^{-3}\,\knorm$. Model\,$\mathcal{B}$ also provided a very good fit to slice\,A at $\chis=719.7/662$ which is comparable to model\,$\mathcal{A}$.

Fig.\,\ref{fig:mcg03_MYTdec_sliceA} shows the best-fit model\,$\mathcal{B}$ to slice\,A and it seems to reproduce quite well the overall spectra. In the following analysis in Section\,\ref{subsec:Time sliced spectral analysis} we will only focus on the \myt model as it assumes a better and a more realistic physical geometry of the reprocessor. In both model\,$\mathcal{A}$ and model\,$\mathcal{B}$ adopted here in slice\,A, the two distinct zones of the wind are ubiquitous, with consistent column densities and outflow velocities (see Table\,\ref{tab:broadband_models_slice1}).

%%%%%%%%%%%%%%%%%%%%%%%%%%%%%%%%% TABLE WITH BROADBAND MODELS %%%%%%%%%%%%%%%%%

\begin{table*}

\begin{tabular}{cc|cc|}

\hline
%\\
Component                                &Parameter                                &Model\,$\mathcal{A}$                 &Model\,$\mathcal{B}$                        \\
                                         &                                         &(Slab reprocessor)                   &(Toroidal reprocessor)                               \\

\hline

\multicolumn{4}{c}{\textbf{Continuum}}\\

\\

\multirow{2}{*}{Primary power-law}

                                         &$\Gamma$                                 &$2.25_{-0.07}^{+0.07}$               &$2.14_{-0.05}^{+0.05}$                  \\

		                                 &norm$^a$                                 &$2.9_{-0.3}^{+0.4}$                  &$2.7_{-0.3}^{+0.3}$                     \\

\\

\multirow{2}{*}{Scattered continuum}

                                         &$\Gamma$                                 &$3.33_{-0.83}^{+0.87}$               &$2.14^{\rm t}$                       \\

					                     &norm$^b$                                 &$1.3_{-0.4}^{+0.4}$                  &$1.3_{-0.2}^{+0.2}$                  \\

\\

\hline

\multicolumn{4}{c}{\textbf{Soft X-ray emission}}\\

\\

\multirow{3}{*}{Photoionized$^c$}

                                         &$\lognh$                                 &$21.5^{*}$                            &$21.5^{*}$                          \\

                                         &$\logxi^d$                               &$1.2^{*}$                             &$1.2^{*}$                           \\

										&$\kappa_{\rm xstar}^e$                   &$1.6_{-0.5}^{+0.4}$                   &$1.8_{-0.2}^{+0.2}$                 \\

\\

\multirow{2}{*}{Thermal}

                                         &kT\,(keV)                                 &$0.79_{-0.05}^{+0.05}$               &$0.78_{-0.05}^{+0.04}$                     \\

										&norm\,($\times10^{-5}$)$^f$              &$1.4_{-0.2}^{+0.2}$                  &$1.5_{-0.2}^{+0.2}$                   \\

\\

\hline

\multicolumn{4}{c}{\textbf{Distant reprocessor and neutral absorber}}\\

\\

\multirow{6}{*}{\texttt{MYTorus}}

                                         &$\Gamma$                                 &--                                  &$2.14^{\rm t}$                          \\

                                         &$\lognh^{g}$                             &--                                  &$23.21_{-0.01}^{+0.01}$                 \\

                                         &$\log (N_{\rm H,S}/\rm cm^{-2})$         &--                                  &--                                      \\

                                         &$\log (N_{\rm H,Z}/\rm cm^{-2})$         &--                                  &--                                      \\

										&$\rm norm_{\rm S}=norm_{\rm L}$          &--                                  &$2.7^{\rm t}$                      \\

		 	             &$\rm norm_{\rm S00}=norm_{\rm L00}=norm_{\rm S90}$       &--                                  &--                               \\

\\

\multirow{2}{*}{\texttt{pexmon}}

                                         &$\Gamma$                                 &$2.25^{\rm t}$                      &--                                       \\

                                         &norm$^h$                                 &$1.3_{-0.5}^{+0.5}$                 &--                                       \\

\\

\multirow{1}{*}{\texttt{zphabs}}

                                         &$\lognh$                                 &$23.38_{-0.02}^{+0.02}$             &--                                      \\

\\

\hline

\multicolumn{4}{c}{\textbf{Highly ionized absorber}}\\

\\

\multirow{3}{*}{	Zone\,1$^k$}

                                         &$\lognh$                                 &$23.2_{-0.1}^{+0.1}$                 &$23.2_{-0.1}^{+0.1}$                   \\

                                         &$\logxi$                                 &$5.4_{-0.2}^{+0.2}$                  &$5.3_{-0.2}^{+0.2}$                    \\

                                         &$\vwc$                                   &$-0.09_{-0.02}^{+0.02}$              &$-0.09_{-0.02}^{+0.02}$                \\

\\

\multirow{3}{*}{Zone\,2$^k$}

                                         &$\lognh$                                 &$24.0_{-0.2}^{+0.3}$                 &$24.1_{-0.2}^{+0.3}$                  \\

                                         &$\logxi$                                 &$6.2_{-1.0}^{+0.4}$                  &$6.2_{-1.0}^{+0.3}$                    \\

                                         &$\vwc$                                   &$-0.35_{-0.02}^{+0.02}$              &$-0.35_{-0.02}^{+0.02}$               \\

\\

%\hline

\\

\multirow{3}{*}{cross-normalization}

                                         &MOS                                      &$0.95\pm0.02$                       &$0.95\pm0.02$                         \\

                                         &FPMA                                     &$1.07\pm0.04$                       &$1.07\pm0.04$                          \\

                                         &FPMB                                     &$1.12\pm0.05$                       &$1.12\pm0.05$                          \\

\\

\\

\hline
      
\\      

Fit statistic                           &$\chi^2/\nu$                            &$714.3/660$                         &$719.7/662$                             \\

\\

\hline

\end{tabular}

\caption{Summary of the broadband best fits parameters applied for slice\,A of \xmmnu (see text for details). $^{\rm t}$ and $^{*}$ denote tied and frozen parameters respectively during fitting.}
\vspace{-5mm}
\begin{threeparttable}
\begin{tablenotes} 
 	\item[a] primary power-law normalization in unit of $\times10^{-3}\,\rm photons\,cm^{-2}\,s^{-1}\,keV^{-1}$,
	\item[b] scattered power-law component in units of $\times10^{-5}\,\rm ph\,cm^{-2}\,s^{-1}\,keV^{-1}$,
	\item[c] \texttt{xstar} emission grid with $\vturb=100\,\rm km\,s^{-1}$, 
    \item[d] ionization parameter with fixed value obtained from the RGS fit,
    \item[e] normalization of the \xstar emission component, in units of $\times10^{-6}$ in terms of $\fcov \frac{L/10^{38}}{D_{\rm kpc}^{2}}$,    
    \item[f] normalization in units of $10^{-5}\,\frac{10^{-14}}{4\upi[D_{A}(1+z)]^2}  \int n_{\rm e} n_{\rm H}dV$, where $n_{\rm e}$ and  $n_{\rm H}$ are the electron and hydrogen densities (measured in $\rm cm^{-3}$) respectively and $D_{A}$ is the angular diameter of the source in $\rm cm$ as defined in the \texttt{mekal} model,	
	\item[g] mean \los column density, integrated over all lines of sight through the torus calculated as $(\upi/4)\nh$,
	\item[h] $(\times10^{-3}\,\rm ph\,cm^{-2}\,s^{-1}\,keV^{-1})$,
    \item[k] \xstar absorption grid with $\vturb=10000\,\rm km\,s^{-1}$.  
\end{tablenotes}
\end{threeparttable}
\label{tab:broadband_models_slice1}
\end{table*}
%\pagebreak

%%%%%%%%%%%%%%%%%%%%%%%%% TABLE ON MYT AND MYTdec for 2 slices %%%%%%%%%%%%%%%%%%%%%%%%%%%%%

\begin{table*}

\begin{tabular}{ccc|cccc|}

\hline
\\
Component                                &&Parameter                       &\multicolumn{2}{c}{Model\,$\mathcal{C}$}                           &\multicolumn{2}{c}{Model\,$\mathcal{C}$}    \\
                                         &&                                &\multicolumn{2}{c}{(decoupled, wind fix)}                                 &\multicolumn{2}{c}{(decoupled, wind vary)}\\

\hline

\multicolumn{7}{c}{\textbf{Continuum}}\\

                                         &&                       &\multicolumn{1}{c}{Slice\,A}   &\multicolumn{1}{c}{Slice\,B}              &\multicolumn{1}{c}{Slice\,A}   &\multicolumn{1}{c}{Slice\,B}\\

\\

\multirow{2}{*}{Primary power-law}

                                        &&$\Gamma$              &$2.29_{-0.07}^{+0.05}$           &$2.29^{\rm t}$                            &$2.36_{-0.06}^{+0.06}$         &$2.36^{\rm t}$\\

		                                &&norm$^a$              &$3.6_{-0.5}^{+0.5}$              &$2.3_{-0.5}^{+0.4}$                       &$4.2_{-0.4}^{+0.5}$            &$3.5_{-0.6}^{+0.8}$\\

\\

\multirow{2}{*}{Scattered continuum}

                                         &&$\Gamma$            &$2.29^{\rm t}$                   &$2.29^{\rm t}$                             &$2.36^{\rm t}$                 &$2.36^{\rm t}$\\

					                    &&norm$^b$            &$1.4_{-0.2}^{+0.2}$               &$1.4^{\rm t}$                             &$1.3_{-0.2}^{+0.2}$             &$1.3^{\rm t}$\\

\\

\hline

\multicolumn{7}{c}{\textbf{Soft X-ray emission}}\\

\\

\multirow{1}{*}{Photoionized emission}

                                         %&&$\lognh^c$                               &$21.5^{\rm f}$           &$21.5^{\rm f}$\\

                                         %&&$\logxi^c$                               &$1.2^{\rm f}$            &$1.2^{\rm f}$\\

		          &&$\kappa_{\rm xstar}$\,($\times10^{-6}$)  &$1.8_{-0.2}^{+0.2}$                &$1.8^{\rm t}$                              &$1.7_{-0.2}^{+0.2}$            &$1.7^{\rm t}$\\

\\

\multirow{2}{*}{Thermal emission}

                                   &&kT\,(keV)                &$0.78_{-0.05}^{+0.05}$           &$0.78^{\rm t}$                               &$0.78_{-0.05}^{+0.05}$        &$0.78^{\rm t}$\\

							      &&norm\,($\times10^{-5}$) &$1.5_{-0.2}^{+0.2}$                &$1.5^{\rm t}$                               &$1.5_{-0.2}^{+0.2}$           &$1.5^{\rm t}$\\

\\

\hline

\multicolumn{7}{c}{\textbf{Distant reprocessor and neutral absorber}}\\

\\

\multirow{4}{*}{\texttt{MYTorus}}

                                   &&$\Gamma$                 &$2.29^{\rm t}$                   &$2.29^{\rm t}$                                &$2.36^{\rm t}$              &$2.36^{\rm t}$\\

                      &&$\log (N_{\rm H,S}/\rm cm^{-2})$     &$24.7_{-0.1}^{+0.4}$              &$24.7^{\rm t}$                                &$>24.6$                     &$>24.6^{\rm t}$\\

                      &&$\log (N_{\rm H,Z}/\rm cm^{-2})$     &$23.16_{-0.02}^{+0.03}$           &$23.34_{-0.05}^{+0.05}$                       &$23.16_{-0.02}^{+0.02}$     &$23.16^{\rm t}$\\

      &&$\rm norm_{\rm S00}=norm_{\rm L00}=norm_{\rm S90}$   &$3.6^{\rm t}$                     &$2.3^{\rm t}$                                 &$4.2^{\rm t}$               &$3.5^{\rm t}$\\

\\

\hline

\multicolumn{7}{c}{\textbf{Highly ionized absorber}}\\

\\

\multirow{3}{*}{\textbf{Zone\,1}}

                     &&$\lognh$                              &$23.1_{-0.1}^{+0.1}$              &$23.1^{\rm t}$                                &\textbf{23.2}$_{-0.1}^{+0.1}$      &\textbf{24.0}$_{-0.1}^{+0.1}$\\

                     &&$\logxi$                              &$4.3_{-0.1}^{+0.1}$               &$4.3^{\rm t}$                                 &$4.1_{-0.1}^{+0.1}$       &$4.1^{\rm t}$\\

                     &&$\vwc$                                &$-0.11_{-0.02}^{+0.01}$           &$-0.11^{\rm t}$                               &$-0.12_{-0.01}^{+0.02}$   &$-0.12^{\rm t}$\\

\\

\multirow{3}{*}{Zone\,2}

                     &&$\lognh$                              &$24.0_{-0.1}^{+0.1}$             &$24.0^{\rm t}$                                 &$23.9_{-0.3}^{+0.2}$     &$23.9^{\rm t}$\\

                     &&$\logxi$                              &$6.2_{-0.9}^{+0.4}$              &$6.2^{\rm t}$                                  &$6.2_{-0.9}^{+0.5}$      &$6.2^{\rm t}$\\

                     &&$\vwc$                                &$-0.36_{-0.02}^{+0.02}$          &$-0.36^{\rm t}$                                &$-0.36_{-0.02}^{+0.02}$  &$-0.36^{\rm t}$\\

\\

\multirow{3}{*}{cross-normalization}

                     &&MOS                                   &$0.95\pm0.02$                    &--                                              &$0.95\pm0.02$           &--\\

                     &&FPMA                                  &$1.05\pm0.04$                    &$0.98_{-0.09}^{+0.11}$                          &$1.07\pm0.04$           &$0.98_{-0.09}^{+0.10}$\\

                     &&FPMB                                  &$1.11_{-0.05}^{+0.04}$           &$1.00_{-0.10}^{+0.12}$                          &$1.12\pm0.04$           &$1.00_{-0.09}^{+0.10}$\\

\\

\hline
      
%\\      

Fit statistic        &&$\chi^2/\nu$                       &\multicolumn{2}{c}{$860.7/761$}                           &\multicolumn{2}{c}{$826.4/761$}    \\

%\\

\hline

\end{tabular}

\caption{Summary of the broadband best fits parameters of model$\mathcal{C}$ applied for slice\,A and B of \xmmnu (see text for details). The important outcome of this result is that the observed spectral variability between the slices can be explained by a drastic increase in $\nh$ by a factor of $\sim10\times$ in zone\,1 of the highly ionized absorber as opposed to a change in the $\nh$ of the neutral absorber. This behavior suggests at a high confidence level ($>99.99\%$) that the observed spectral variability is caused by the highly ionized material rather than a neutral inhomogeneous absorber. All the parameters and units are the same as in Table\,\ref{tab:broadband_models_slice1}.}
\label{tab:broadband_models_2slices}

\end{table*}

\subsection{Time sliced spectral analysis}
\label{subsec:Time sliced spectral analysis}

\begin{figure}
\includegraphics[scale=0.5]{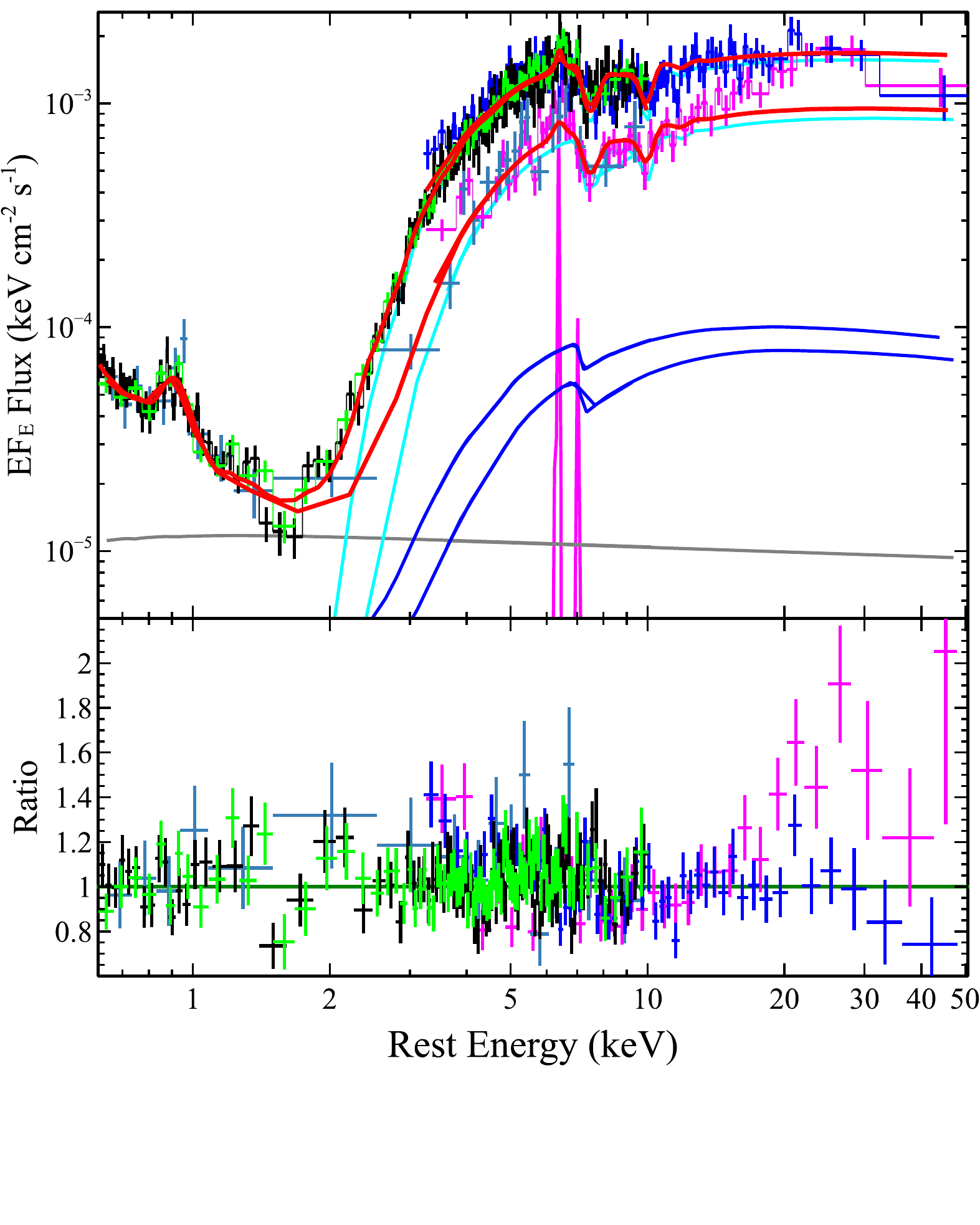}
  \vspace{-15mm}
\caption{Top panel: Best-fit model\,$\mathcal{B}$ superimposed (red) between slice\,A and B. The corresponding \myt model contributions are (zeroth-order) transmitted absorbed primary component (cyan), the reflected continuum (blue) and the \feka and \fekb fluorescent line emission lines (magenta). The distant scattered component is shown in gray. Bottom panel: The corresponding data/model ratio which show a clear excess in the residuals above 20\,keV in slice\,B, which cannot be accounted for in the coupled \myt model. The separate thermal and photoionized components are not included in the plot for clarity (but are included in the model). Note that for plotting purposes we adopted the combined \nustar FPMA$+$FPMB spectra.}
	\label{fig:mcg03_MYT_sliceAB_wind_novary}
\end{figure}

From the above broadband spectral analysis, we show that both model\,$\mathcal{A}$ and $\mathcal{B}$ successfully fitted slice\,A. However, B18 detected a rapid spectral variability in \mcg between slice\,A and slice\,B which was likely caused by an obscuration event during the \xmmnu observation. In the following we verify whether, by adopting a more self-consistent model for the neutral absorber such as \myt (model\,$\mathcal{B}$), rather than a slab reprocessor (model\,$\mathcal{A}$), the result discussed in B18 is still valid. In particular, we explore whether the obscuration event is caused by: (i) a transiting neutral absorber within a clumpy torus or rather is (ii) the result of the inhomogeneous nature of the highly ionized and fast disc-wind where a filament or clump rapidly crosses the \los (as also seen in \pds \citealt{Matzeu16}). We first attempt to test scenario\,(i) with model\,$\mathcal{B}$ by only varying the \los column density and the primary continuum normalization, whilst keeping the wind parameters tied between the two slices. However such model simply fails to fully reproduce the curvature in slice\,B, leaving a strong excess particularly between $20$--$40\,\kev$ and yields a poor fit ($\chis=893.0/762$) as shown in \fig\,\ref{fig:mcg03_MYT_sliceAB_wind_novary}. The harder shape of the spectrum in slice\,B due to a higher obscuration compared to slice\,A.

We then explored a more complex geometry of the reprocessor by adopting the `decoupled' configuration of \myt defined as model\,$\mathcal{C}$. By following the methodology presented in details in \citet{Yaqoob12}, we want to represent a physical scenario where the neutral absorber is inhomogeneous in nature and hence characterized by a more `patchy' distribution of reprocessing clouds. This can be achieved by decoupling the inclination angle parameters of the zeroth-order (\los) and reflected continua and allowing the corresponding column density defined as $N_{\rm H,Z}$ (\los $\nh$) and $N_{\rm H,S}$ (global $\nh$) respectively. The inclination of the zeroth-order component is fixed at $90^{\circ}$ whereas the reflected component at $0^{\circ}$. The fluorescence emission lines, and the relative normalizations are all tied to the primary power-law i.e., $\rm norm_{\rm S00}=norm_{\rm L00}=norm_{\rm S90}=norm_{\rm intr\,zpowl}$. As in model\,$\mathcal{B}$, all the corresponding constant factors $\rm A_{\rm S00}=A_{\rm L00}=A_{\rm S90}$ are set to unity.

\begin{figure*}
\centering
%\begin{subfigure}[t]{0.3\textwidth}
%\centering
\includegraphics[width=\textwidth]{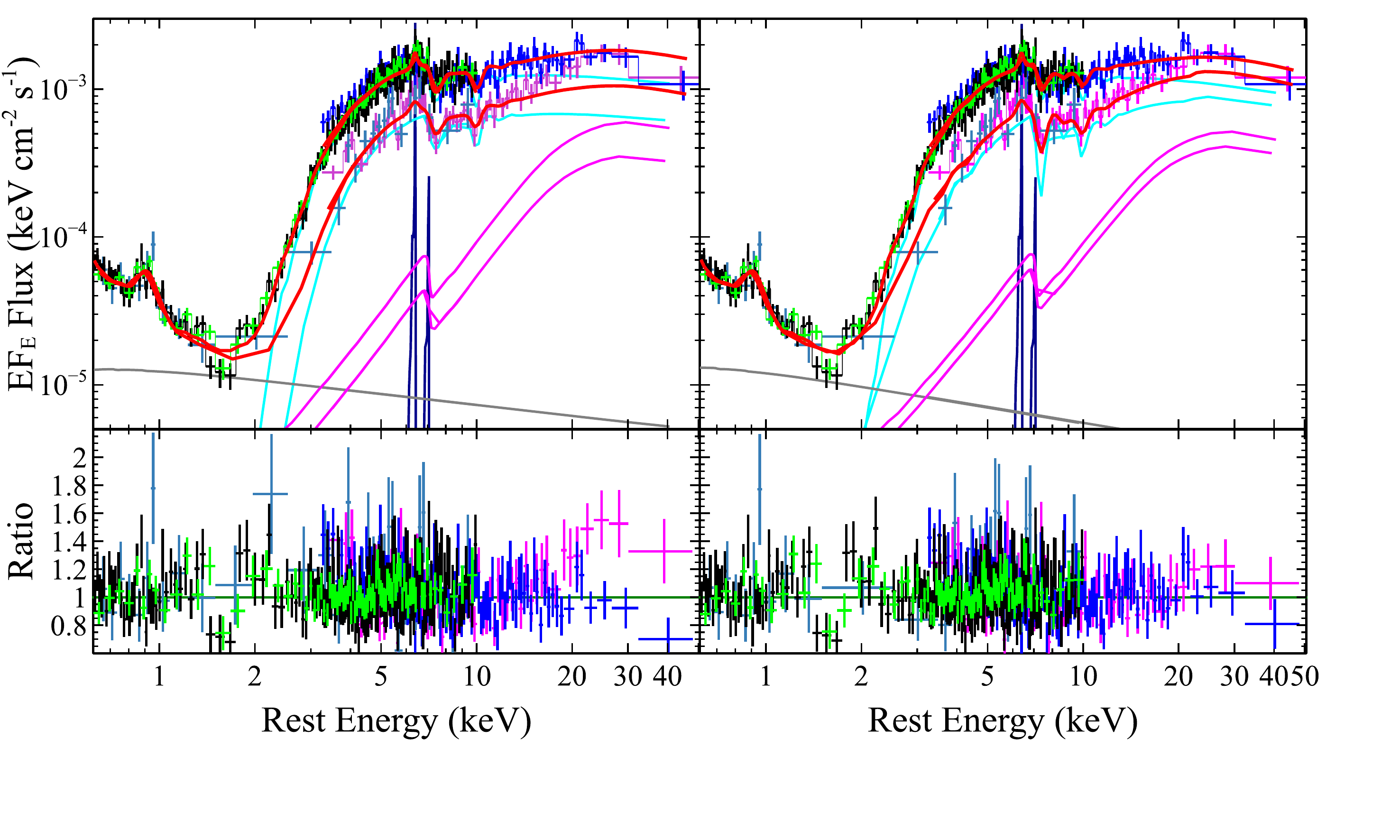} 
%\caption{Generic} \label{fig:timing1}
%\end{subfigure}
\vspace{-1cm}
 \caption{Best-fit \myt decoupled mode (model\,${C}$), applied to slice\,A and slice\,B where the left panel shows a variable transmitted $N_{\rm H,Z}$ (and constant wind), whereas the right panel shows a constant transmitted $N_{\rm H,Z}$ (and variable wind). Only the latter is able to fully account for the pronounced spectral curvature seen up to $40\,\kev$ in slice\,B. Note that for plotting purposes we adopted the combined \nustar FPMA$+$FPMB spectra. As the soft X-ray model components such as \mekal and \xstar emissions are not variable and do not affect the \myt parameters, we omitted them here for clarity. The distant scattered component is shown in gray and the \myt components are defined as in \fig\,\ref{fig:mcg03_MYT_sliceAB_wind_novary}.}
\label{fig:MYTdec_2sl_C1_2}
\end{figure*}

In this model we find that the global $\nh$ is much larger than the zeroth-order at $\log(N_{\rm H,S}/\rm cm^{-2})=24.7_{-0.1}^{+0.4}$ which imprints a much stronger reflection component in the spectrum. Although better than the coupled configuration, model\,$\mathcal{C}$ still forces a $\sim40\%$ decrease of the primary continuum which still produces some excess residuals in slice\,B between $20$--$40\,\kev$ (see bottom-left panel in Fig.\,\ref{fig:MYTdec_2sl_C1_2}). Although the transmitted $\nh$ increases and hence explain well the spectral curvature in slice\,B, is still not enough to fully account for it (see top-left panels in Fig.\,\ref{fig:MYTdec_2sl_C1_2}). Instead, this seems to suggest that an additional variable absorber, such as the wind, is required to explain the increase in hardness of slice\,B. In summary by investigating case\,(i) with model\,$\mathcal{C}$, resulted in an improved fit compared to model\,$\mathcal{B}$ at $ \chis=860.7/761$ (corresponding to a $>99.99\%$ improvement). Even if the fit statistically improves, model\,$\mathcal{C}$ is not able to fully reproduce the spectral shape of slice\,B, leaving strong residuals in the $20$--$40\,\kev$ range. Subsequently we tested case\,(ii) with model\,$\mathcal{C}$ by keeping the transmitted $N_{\rm H,Z}$ of the \myt model tied between slices, but allowing the column density of the slower ($\vwc=0.1$) zone to vary. In Fig.\,\ref{fig:MYTdec_2sl_C1_2} (right panels) we show the best-fit model\,$\mathcal{C}$ for case\,(ii) overlaid on the fluxed slice\,A and B spectra (top) and the corresponding residuals (bottom). This time the spectral curvature in slice\,B is well reproduced. This is explained by a drastic increase in column density of the slower ionized absorber (zone\,1) by almost one order of magnitude i.e., from $\lognh=23.2\pm0.1$ to $\lognh=24.0\pm0.1$ between slice\,A and B. 

In summary, the observed variability of spectral hardness and \los column density is best modelled by a change in $\nh$ of the ionized absorber rather than by neutral absorption. Moreover, as in B18, we also found that there is no improvement in the fit by letting the $\nh$ of the fast zone\,2 to vary, hence suggesting that the density change in the $\vwc=0.1$ zone is what drives the observed variability. We also found that the overall power-law normalization does not decrease as drastically ($\sim10\%$) and the model now converges at $\sim20\,\kev$, hence all the variations occur at lower energies due to absorption variability. Thus this model led to an excellent fit to the overall data at $\chis=826.4/761$, providing a substantial improvement of $\dchis=34.3/1$ (i.e., $\sim6\sigma$).

%\pagebreak

%\pagebreak
\section{Discussion and conclusion}
\label{sec:Discussion and conclusion}

We presented a detailed broadband spectral analysis of the X-ray spectra emission within $0.6$--$50\,\kev$ of \mcg, where we successfully de-convolved, by using \myt, all the layers of absorption. We infer a complex structure of absorbers consisting of: (a) a primary power-law component absorbed by a fully covering neutral medium likely associated with the inhomogeneous toroidal absorber, which is thicker in the equatorial plane, (b) a reflected component from a distant reprocessor, and (c) a multi-phase highly ionized fast outflow. In addition to this, we considered the analysis of the RGS data which allowed us to better understand the origin of the soft X-ray emission. Our main findings are summarized and discussed in the following.

\subsection{The nature of the soft X-ray emission}
\label{subsec:The nature of the soft X-ray emission}

The soft X-ray emission of \mcg can be well described with a superposition of a power-law scattered component and several narrow emission lines. The latter are associated with optically thin, photoionized gas as well as a weaker collisionally ionized plasma component consistent with starburst activities in the host galaxy. From the luminosity measured in the IR band, \citet{Oi10} were able to robustly estimate the star formation rate (SFR) in \mcg to be $\rm SFR=9.4\,\Msun\,yr^{-1}$. From the SFR and the $L_{\rm x}$--SFR  correlations derived from various sample of LIRG and ULIRG \citep[e.g.,][]{Ranalli03,PereiraSantaella11,Mineo14}, we derived the predicted soft X-ray luminosity to be $L_{(0.5-2)\,\kev}=3\times10^{40}\,\ergs$ which is consistent with what it is measured in \mcg for the star burst component. In terms of the photoionized component, we measured a soft X-ray luminosity of $L_{(0.5-2)\,\kev}\sim10^{41}\,\ergs$.

We did not observe any variability in the soft band in respect to the 2010 \suzaku observation (see B18). All the emission lines are unresolved at the RGS spectral resolution ($\sigma_{\rm v}\lesssim1700\,\kms$). From the results outlined in Section\,\ref{subsec:RGS analysis}, the O\,\textsc{vii} triplet ratio gave us an upper limit on the gas density of $n_{\rm e}<10^{9}\,\rm cm^{-3}$ whereas from the photoionization modelling we obtained an ionization state of $\xi\sim10\,\rm erg\,cm\,s^{-1}$. Given the ionizing luminosity of $\lion\sim10^{45}\,\ergs$ in \mcg, this places a lower limit on the radial distance of the gas of $R>0.1\,\rm pc$. On the other hand a subsequent \chandra imaging observation of \mcg in 2016 reveal that the soft X-ray emitter is largely point-like, within an arc second radius of the nucleus (Braito et al. in prep). For \mcg this corresponds to an emitting gas that is confined within a scale of few hundreds of parsec and although this distance scale is poorly constrained, it is likely constant with NLR gas on approximately parsec scales or greater as often seen in other Seyfert\,2 galaxies \citep[e.g.,][]{Sako00,Kinkhabwala02,Braito17}. As shown in Fig\,\ref{fig:rgs_xstar_plot}, the dominant contributor to the emission lines is mostly a distant gas photoionized by the AGN often observed in Seyfert\,2 galaxies \citep[e.g.,][]{GuainazziBianchi07}. However in all the adopted models we also observed a contribution from a weak thermal emission component with temperature of $kT\sim0.8\,\kev$. Note that if we did not include the photoionized emission, the luminosity of the emission lines component (parameterized with \mekal) would be simply too high with respect to the soft X-ray luminosity expected from the star formation rate.

\subsection{X-ray broadband variability: clumpy torus or highly ionized disc-wind}
\label{subsec:X-ray broadband variability}

The variability in the \xmmnu spectra in \mcg was caused by a rapid eclipsing event as discussed in B18. In this work we further investigated the cause of this event. We initially tested a scenario where the variability between the slices (defined in Section\,\ref{subsec:Description of the xmmnu lightcurves}) could be explained solely with a change in $\nh$ of the neutral absorber and the primary continuum normalization by adopting the coupled \myt configuration whilst keeping the disc-wind \xstar parameters fixed between the slices. As discussed in Section\,\ref{subsec:Time sliced spectral analysis}, in the coupled \myt configuration the column densities of the transmitted and reflected components are coupled together (i.e., have the same $\nh$). We found that in this scenario the $\nh$ only increased by $\sim30\%$, while the primary continuum decreased by $\sim50\%$. Moreover, as there is no spectral variability between the slices $>20\,\kev$, the excessive drop in normalization fails to account for the sharp hardening in the spectrum in slice\,B. This results in a very prominent excess residuals in the \nustar spectrum above $20\,\kev$ in slice\,B as shown in Fig.\,\ref{fig:mcg03_MYT_sliceAB_wind_novary}. 

In B18, it was tested whether the spectral variability could be explained by the presence of a classical clumpy neutral absorber. Although it produced an acceptable fit, this scenario was also ruled out due to its excessively high column density at $\lognh>24.5$ which required $\sim99\%$ covering (during slice\,B) and the unphysically high luminosity derived when Compton (down)scattering effects were taken to account for such high $\nh$. In this work, we attempted to test the above scenario with the more physical \myt model by decoupling the $\nh$ of the out of \los reflected and zeroth-order (transmitted) components (model\,$\mathcal{C}$). The patchy distribution of the reprocessor assumed in model\,$\mathcal{C}$, allowed us to better account for the spectral curvature, in slice\,B (see Section\,\ref{subsubsec:xmmnu modeling with MYTorus}). Despite this, the overall normalization of the primary continuum still excessively decreased and hence significant residuals were still present $\gtrsim20\,\kev$ in slice\,B (see left panel of Fig\,\ref{fig:MYTdec_2sl_C1_2}).

When fitting the \xmmnu data with the \myt decoupled mode, we were able to measure the column densities of both the out of the \los and transmitted reprocessors. Remarkably, we found that the measured global $N_{\rm H,S}$ was indeed much larger that the \los $N_{\rm H,Z}$ by at least an order of magnitude (see Table\,\ref{tab:broadband_models_2slices}). Such an extreme difference between these two absorbers can be geometrically explained with an overall patchy toroidal reprocessor which is broadly Compton-thin with a relatively small equatorial thick layer out of the \los as shown schematically in in \fig\ref{fig:mcg03_schema_v8}.

\mcg is not an outlier in this respect, in fact the difference between the global and zeroth-order $\nh$ which suggest an inhomogeneous torus has been already observed in other Seyfert\,2 galaxies such as e.g., NGC\,4945 \citep{Yaqoob12}, Markarian\,3 \citep{Yaqoob15}. This sort of inhomogeneous nature of the torus has been widely accepted in the scientific community by developing torus models considering a clumpy gas distribution \citep[e.g.,][]{Elitzur06,Nenkova08a,Nenkova08b}.  Furthermore in the most recent models, the overall absorption can be quantified more as a viewing probability dependent on the physical properties of the cloud i.e., size and location, which typically tend to be distributed towards the equatorial plane.

Our modelling with the self-consistent model for a toroidal absorber confirms that the main driver for the observed variability is the wind rather than a clumpy neutral absorber. In all the models tested, when the wind column is held constant between the slices, the models require the normalization of the primary continuum to drop in slice\,B to compensate, leaving an excess above $20\,\kev$ in the residuals. However this can be ruled out, as such a drop in the continuum is not seen in the $20$--$40\,\kev$ lightcurve. Instead, if the wind $\nh$ is allowed to increase in slice\,B, then this can naturally account for the increase in hardness of slice B, while the overall continuum normalization is now constant, consistent with the lack of variability above $20\,\kev$. Thus this result suggested at high confidence level $>99.99\%$ that the eclipsing event was not caused by a neutral absorber in the clumpy torus but instead driven by a transiting clump (or filament) in an inhomogeneous and highly ionized disc-wind located a few hundreds of $\rg$ from the black hole (B18). Indeed such rapid absorption event caused by the highly ionized disc-wind was also observed in \pds \citep{Gofford14,Matzeu16}, which is considered to be hosting the prototype disc-wind \citep{Reeves18b}, as well as e.g., in PG\,1211$+$143 \citep{Pounds03}, PG\,1126--041 \citep{Giustini11} and APM\,08279$+$5255 \citep{Chartas03}. Despite the lower X-ray luminosity that characterizes \mcg, the presence of such powerful and highly variable disc-wind resembles the one observed in the more powerful quasars with the addition that they are not usually observed in standard obscured Seyfert galaxies. The presence of variable ionized inhomogeneous outflows have also been observed in the UV band, located at larger distance $\sim0.01$--$10\,\rm pc$ \citep[e.g.,][]{Capellupo13,McGraw18}, which are increasingly supporting the multi-phase structure of these winds. For example in the narrow-line Seyfert\,1 galaxy WPVS\,007, \citet{Leighly15} observed for the first time an occultation event in the UV associated with broad absorption line (BAL) gas in the torus; whereas in the Seyfert\,1 galaxy NGC\,5548 an ongoing long-term obscuration is observed in both the X-rays and UV bands \citep{Kaastra14}. Interestingly, \citet{Hamann18} detected a fast UV counterpart of the X-ray wind in \pds which was measured at a comparable outflow velocity of $\vw\sim0.3c$.

\subsection{Physical Properties and location of the clumpy disc-wind}

Now we investigate the main properties of the highly ionized fast disc-wind and compare them with the results obtained in B18. From the observed outflow velocity ($\vw$) inferred from the two blueshifted \fe absorption features, we can estimate a lower limit on the launching radius. This can be done by equating the $\vw$ with its escape radius, thus $R_{\rm esc}=(2c^{2}/\vw^{2})\rg$ where $\rg=G\mbh/c^2$ is the gravitational radius which corresponds to $1\rg\sim1.5\times10^{13}\,\rm cm$ for a black hole mass of $\mbh\sim10^{8}\Msun$ in \mcg (B18). We therefore derive $R_{\rm in,1}\sim140\,\rg\,(\sim2\times10^{15}\,\rm cm)$ and $R_{\rm in,2}\sim15\,\rg\,(\sim2\times10^{14}\,\rm cm)$ for zone\,1 ($\vwc=-0.12_{-0.01}^{+0.02}$) and zone\,2 ($\vwc=-0.36\pm0.02$) respectively. These results suggest that we are viewing the disc-wind through a \los that intercepts two distinct streamlines launched at different radii as also observed in \pds \citep{Reeves18b}. We then utilize these measurements to estimate the disc-wind energetics by quantifying the mass outflow rate ($\mw$) expressed as $\mw=f\mu\upi m_{\rm p}  \vw \nh \rm R_{\rm in}$, derived by \citet{Krongold07}, where a biconical geometry of the outflow is assumed. The constant factor for cosmic elemental abundances is set to $\mu=n_{\rm H}/n_{\rm e}=1.2$ and $R_{\rm in}$ is the inner radius or the starting point of the disc-wind. The function $f$ takes into account the inclination angle with respect to the \los and the disc. Unlike in \pds, where the disc-wind's solid angle was directly measured \citep{Nardini15}, the geometry of the system in \mcg is currently uncertain and hence we adopt a face value of $f=1.5$ (see Appendix 2 in \citealt{Krongold07}).

Thus for the slow zone\,1, observed in slice\,A, we obtain a mass outflow rate of $\dot M_{\rm w,z1}\sim1.2\times10^{25}\,\rm g\,s^{-1}\,(\sim0.3\Msun\,yr^{-1})$ which implies a kinetic power of $\dot E_{\rm w,z1}\sim8\times10^{43}\,\ergs$. This corresponds to $\sim3\%$ of the bolometric luminosity and hence falls within the, theoretically predicted, minimum requirement i.e., $\dot E_{\rm w}/\lbol\sim0.5$--$5\%$ for providing a feedback mechanisms between the central SMBH and its host galaxy \citep[e.g.,][]{DiMatteo05,HopkinsElvis10}. This result is consistent with B18 with the only difference being that a constant $\vw$ is assumed across the modelling. On the other hand in slice\,B, the drastic increase in $\nh$ results in a considerably higher kinetic power of $\dot E_{\rm w,z1}\sim5\times10^{44}\,\ergs$ which corresponds to $\sim15\%$ of $\lbol$ and hence exceeding the typical values quoted by AGN feedback models. By assuming that the outflow velocity is comparable to the Keplerian velocity across the source i.e., $v_{\rm K}\sim\vw=0.12c$ and from the measured timescale of this obscuration event $\Delta t=120\,\ks$ we derive the radial size of the absorber to be $\Delta R_{\rm c}=v_{\rm K}\Delta t\sim30\rg\,(\sim4\times10^{14}\,\rm cm)$. The hydrogen number density of the cloud is defined as:

\begin{equation}
n_{\rm H}\sim\frac{\Delta N_{\rm H}}{\Delta R_{c}}\sim\frac{\Delta N_{\rm H}r_{\rm g}^{1/2}}{\sqrt{2}c \Delta t},
\end{equation}

\noindent where $\Delta \nh$ is the observed change in column density between slice\,A and B and $r_{\rm g}=R/\rg$ is the radial distance in units of $\rg$. From the definition of the ionization parameter i.e., $\lion/n_{\rm H}R^{2}$ \citep{Tarter69} we have $n_{\rm H}\sim n_{\rm e}=\lion/\xi R^{2}$ and hence equating these densities we get:

\begin{equation}
 r_{\rm g}^{5/2}=\frac{L_{\rm ion}}{\xi}\frac{\sqrt{2} \Delta t\,c^5}{\Delta N_{\rm H}}\left (GM_{\rm BH} \right )^{-2}.
 \end{equation} 

\noindent From an ionization of $\xi\sim13000\,\rm erg\,cm\,s^{-1}$, an observed change in column density of $\Delta \nh\sim8.4\times10^{23}\,\rm cm^{-2}$ (obtained in model\,$\mathcal{C}$) and by assuming an ionizing luminosity of order of $\lion\sim10^{45}\, \ergs$, the location of the eclipsing cloud is derived to be $R\sim340\,\rg\,(5\times10^{15}\,\rm cm)$, which is consistent to what was measured in B18. These results suggest that we are viewing through a clumpy disc-wind at a typical distance of few hundreds of $\rg$ from the central black hole, launched between tens to hundreds of $\rg$. Regarding the fast zone\,2, the energetic that can be derived are purely speculative as, already noted in B18, the $\nh$ can only be constrained by a given best-fit ionization value as is otherwise degenerate with the ionization(see Section\,\ref{subsec:Analysis of slice A}).

\begin{figure*}
\centering
%\begin{subfigure}[t]{0.3\textwidth}
%\centering
\includegraphics[width=\textwidth]{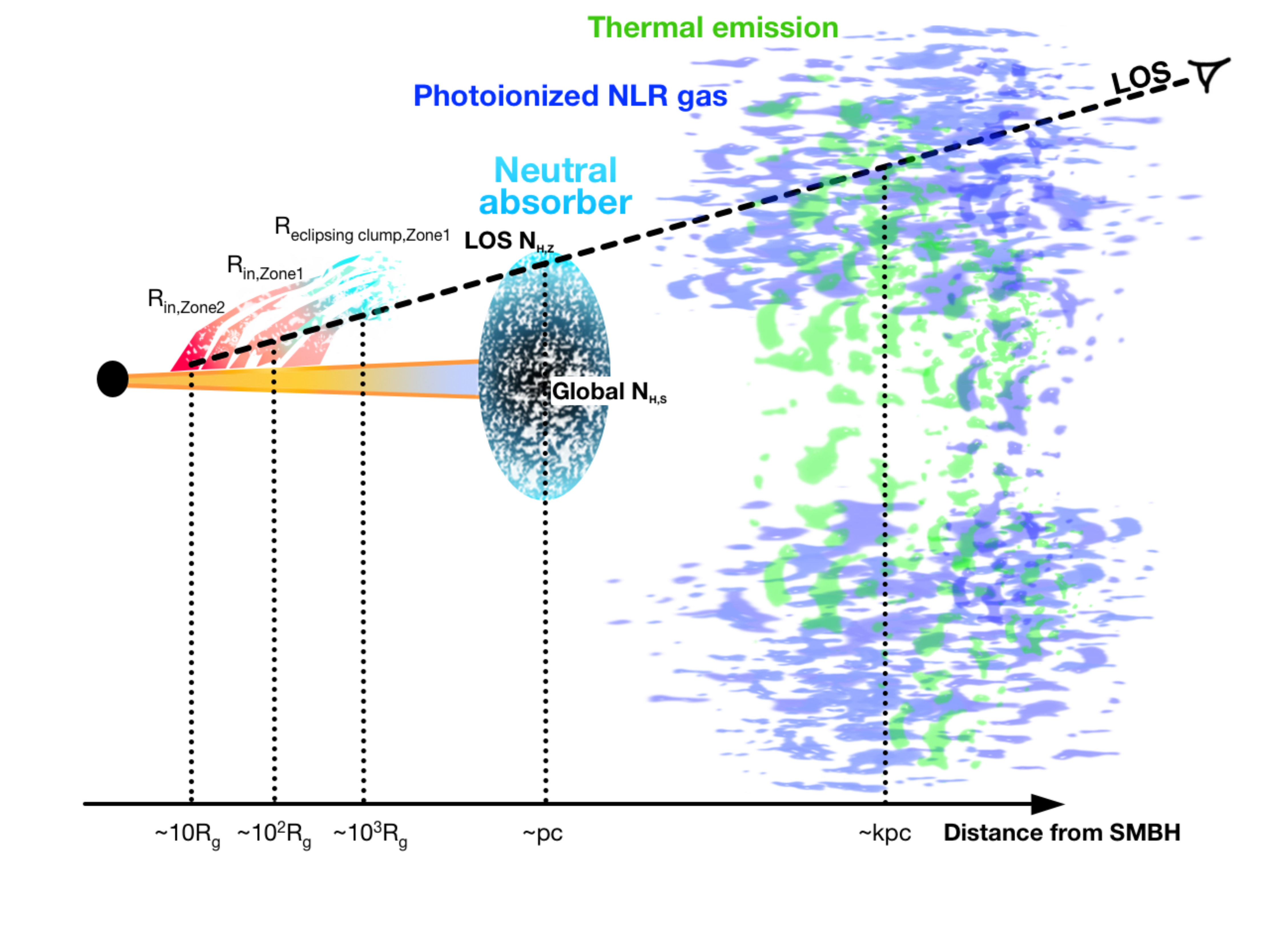} 
%\caption{Generic} \label{fig:timing1}
%\end{subfigure}
\vspace{-1cm}
 \caption{Schematic representation of the possible geometry for \mcg. Our \los is represented by the dashed line. The different components of the system are shown with the corresponding distance scales from the SMBH. Here there are three main regions i.e., the highly ionized outflow, the clumpy torus and the diffuse NLR gas are illustrated. The highly ionized outflow is located closer in, where the regions from high to lower ionization are shown in red to cyan respectively. The launch radii and location of the eclipsing ionized cloud of outflowing material are reported. The clumpy torus is located at intermediate scales where the two distinct column densities (measured in the \myt decoupled mode) of the embedded neutral material are indicated. The global $N_{\rm H,S}$ is possibly associated with Compton-thick material might be distributed equatorially. On the other hand, the measured column density that crosses the \los ($N_{\rm H,Z}$) is associated with Compton-thin material that may be located at  the edge of the clumpy torus. The diffuse gas is located at larger scales where the photoionized and thermal emitters are illustrated in blue and green respectively.}
\label{fig:mcg03_schema_v8}
\end{figure*}

A physical picture of the possible geometry of such a system is schematically illustrated in \fig\ref{fig:mcg03_schema_v8} where the estimated distances of the launching radii of both zones and location of the eclipsing clump and distant circumnuclear gas are illustrated. From the overall information that it is gathered in this work, we have a scenario in which there are three main regions i.e., the highly ionized inhomogeneous fast outflows, the clumpy toroidal neutral absorbers and the diffuse circumnuclear NLR scale gas. The highly ionized outflow is likely inhomogeneous in structure and it is located closer in to the SMBH, where the regions from high to lower ionization are shown in red to cyan respectively. The clumpy torus is located at intermediate scales where we were able to measure two distinct column densities (in model\,$\mathcal{C}$) of the embedded neutral material (light to dark blue represent the increasing in $\nh$). 

The global $N_{\rm H,S}$ can be associated with a compact region of Compton-thick material that might be distributed equatorially out of our \los. On the other hand, the measured column density that intercept the \los ($N_{\rm H,Z}$) might be associated with Compton-thin material that is located at the edge of the clumpy torus. The diffuse circumnuclear gas is situated at larger scales ($>$\,BLR) where the photoionized and thermal emitters are illustrated in blue and green respectively. The low covering factor that characterizes this gas might suggest that it is also inhomogeneous in structure where the dominant component can be associated with photoionized emission caused by the AGN. Although \mcg is classified as Seyfert\,2 galaxy the fortuitous \los (black dash line) through Compton-thin material. Thanks to the synergy between \xmmnu X-ray observatories, it was possible to build a complete picture of this complex system.

%\pagebreak

%\pagebreak
\section*{acknowledgements}

We want to thank the referee for the detailed and helpful report that improved the clarity of this paper. GAM also thank Dr Emanuele Nardini for the useful discussions. GAM and MLP are supported by European Space Agency (ESA) Research Fellowships. Based on observations obtained with \xmm, an ESA science mission with instruments and contributions directly funded by ESA Member States and NASA. GAM, VB, PS, AC and RDC acknowledge support from the Italian Space Agency (contracts ASI-INAF  I/037/12/0 and ASI-INAF n.2017-14-H.0). JR acknowledges financial support through grants NNX17AC38G, NNX17AD56G and HST-GO-14477.001-A. C.C. acknowledges funding from the European Union's Horizon 2020 research and innovation programme under the Marie Sk{\l}odowska-Curie grant agreement No. 664931. %This paper is dedicated to my dear friend and fellow student Clint Clements that passed away early this year and you are truly missed.

%\pagebreak

%We thank the anonymous referee for their helpful report. GM and VB acknowledge support from the Italian Space Agency (ASI INAF NuSTAR I/037/12/0). 

\bibliographystyle{mn2e}
\bibliography{matzeu_references}

\end{document}